\documentclass[fleqn,usenatbib]{mnras}

\usepackage{newtxtext,newtxmath}
\usepackage{booktabs} 
\usepackage{colortbl}
\usepackage[table,xcdraw]{xcolor}
\usepackage{makecell}
\usepackage{gensymb}
\usepackage[T1]{fontenc}

\DeclareRobustCommand{\VAN}[3]{#2}
\let\VANthebibliography\thebibliography
\def\thebibliography{\DeclareRobustCommand{\VAN}[3]{##3}\VANthebibliography}

\usepackage{graphicx}	% Including figure files

\usepackage{amsmath}% Advanced maths commands
\newcommand{\RM}{{\mathrm{RM}}}

\newcommand{\RMa}{\langle\RM\rangle}
\newcommand{\DM}{{\mathrm{DM}}}
\newcommand{\DMa}{\langle\DM\rangle}

\newcommand{\Bpa}{{\langle B_{\parallel} \rangle}}
\newcommand{\Bp}{{B_{\parallel}}}

\newcommand{\B}{\vec{B}}
\newcommand{\Bm}{{\vec{B}}}
\newcommand{\Bf}{\vec{b}}
\newcommand{\lb}{{\ell_{b}}}
\newcommand{\co}{{c_{0}}}

\newcommand{\lne}{\ell_{\delta \ne}}
\newcommand{\gl}{\mathrm{Gl}}
\newcommand{\gb}{\mathrm{Gb}}
\newcommand{\Ca}{{\mathcal{C}_{1}}}
\newcommand{\Cb}{{\mathcal{C}_{2}}}
\newcommand{\Cc}{{\mathcal{C}_{3}}}
\newcommand{\Co}{{\mathcal{C}_{0}}}
\renewcommand{\ne}{n_{\rm e}}
\newcommand{\nea}{\langle \ne \rangle}
\renewcommand{\vec}[1]{\boldsymbol{#1}}	
\renewcommand{\L}{\mathrm{L}}
\renewcommand{\S}{\mathcal{S}}
\newcommand{\muG}{{\mu\mathrm{G}}}
\newcommand{\de}{^{\circ}}
\newcommand{\rad}{{\mathrm{rad}}}
\newcommand{\m}{{\mathrm{m}}}
\newcommand{\pc}{{\mathrm{pc}}}
\newcommand{\cm}{{\mathrm{cm}}}
\newcommand{\dd}{{\mathrm{d}}}
\newcommand{\kpc}{{\mathrm{kpc}}}
\newcommand{\ndata}{{\rm N}}
\newcommand{\dista}{\mathrm{\L_{Ind}}}
\newcommand{\distdm}{\mathrm{\L}_{\DM}}

\newcommand{\kurt}{{\mathcal{K}}}
\renewcommand{\varsigma}{\sigma}
\newcommand{\redchi}{{\chi^{2}}}

\newcommand\Eq[1]{equation~\eqref{#1}}
\newcommand\Fig[1]{Fig.~\ref{#1}}
\newcommand\Sec[1]{Sec.~\ref{#1}}
\newcommand\Tab[1]{Table~\ref{#1}}
\newcommand\App[1]{Appendix~\ref{#1}}

\graphicspath{{./}{figs/}}

\newcommand\rev[1]{{#1}}
\newcommand\revb[1]{{#1}}
\usepackage{soul}
\newcommand\revst[1]{{}}
\title[Milky Way's magneto-ionic medium]{Probing the magneto-ionic medium of the Milky Way using pulsars}

\author[Dhakal and Seta]{
Saakshi Dhakal$^{1,2}$\thanks{E-mail: \href{mailto:u7120811@anu.edu.au}{u7120811@anu.edu.au}, \href{mailto:saakshidhakal46@gmail.com}{saakshidhakal46@gmail.com}} and
Amit Seta$^{2}$\thanks{E-mail: \href{mailto:amit.seta@anu.edu.au}{amit.seta@anu.edu.au}}
\\
$^1$Research School of Earth Sciences,  Australian National University, Canberra, ACT 2601, Australia\\
$^2$Research School of Astronomy and Astrophysics,  Australian National University, Canberra, ACT 2611, Australia\\
}

\date{Accepted XXX. Received YYY; in original form ZZZ}

\pubyear{2024}

\begin{document}
\label{firstpage}
\pagerange{\pageref{firstpage}--\pageref{lastpage}}
\maketitle

\begin{abstract}
Magnetic fields are fundamental to the dynamics of the interstellar medium (ISM) in spiral galaxies and are often separated into large-scale, regular ($\boldsymbol{B}$) and small-scale, random ($\boldsymbol{b}$) components. The thermal electron density, $n_{\rm e}$, can also be divided into large-scale, diffuse, $\langle n_{\rm e} \rangle$, and small-scale, clumpy, $\delta n_{\rm e}$, components. Estimating the properties of $b$ and $\delta n_{\rm e}$ from observations, even within the Milky Way, has long been challenging. This work addresses the challenge using pulsars, which probe the Milky Way's magneto-ionic medium. Using data of more than 1200 pulsars from the Australia Telescope National Facility pulsar catalogue, we combine dispersion (${\rm DM}$) and rotation (${\rm RM}$) measures with theoretical models to estimate both small- and large-scale properties of the Galactic magnetic field and thermal electron density. We find no significant correlation between the average parallel magnetic field strength, $\langle B_{\parallel} \rangle [\mu{\rm G}] = 1.232\,{\rm RM}\,[{\rm rad\,m^{-2}}]/{\rm DM}\,[{\rm pc\,cm^{-3}}]$, and pulsar distance. For pulsars within $20\,{\rm kpc}$, we estimate $|B| \approx 1.2\,\mu{\rm G}$ and $\langle n_{\rm e} \rangle \approx 0.05\,{\rm cm}^{-3}$. More importantly, we determine correlation lengths of small-scale components, $\ell_{b} \approx 20$ -- $30\,{\rm pc}$ and $\ell_{\delta n_{\rm e}} \approx 250$ -- $300\,{\rm pc}$. At smaller distances, $B$ remains roughly constant, while $\langle n_{\rm e} \rangle$ increases and both length scales decrease. These results refine our understanding of fundamental scales in the magneto-ionic medium, aiding the interpretation of extragalactic ${\rm RM}$s and providing insights into the role of magnetic fields in galaxies.
\end{abstract}


\section{Introduction} \label{sec:intro}

Understanding the magnetic fields in our Milky Way is essential for uncovering the astrophysical processes that shape its evolution. These fields play a pivotal role in regulating star formation \citep{Krumholz2019, PattleEA2023}, guiding cosmic ray propagation \citep{ShukurovEA2017, RuszkowskiP2023}, and driving the dynamics of the interstellar gas \citep{ShettyO2006, PlanckXXXV_2016}.

Several observational probes are used to trace the magnetic fields within the interstellar medium (ISM) of galaxies: optical polarisation, Zeeman splitting of spectral lines, polarised emission from dust and molecules, polarised synchrotron emission, and Faraday rotation \citep{KleinF2015}. In general, each probe is useful for probing different regions or phases of the ISM. For example, Faraday rotation probes the magnetic field in the diffuse, ionised ISM, while dust polarisation or Zeeman effect probes the colder, denser ISM \citep{Ferriere2020, Martin-AlvarezEA2024, SetaMCG2025}. 

Observationally, magnetic fields in spiral galaxies are divided into large-scale, regular (on scales of a few $\kpc$) and small-scale, random (on scales of $\lesssim 100\,\pc$) components \citep{BeckEA1996, BrandenburgS2005, Beck2016, Rincon2019, ShukurovS2021}. Similarly, the thermal electron density is also divided into large-scale, diffuse (on scales of $1$ \text{--} $2\,\kpc$) and small-scale, clumpy (on scales of a $10$ -- $100\,\pc$, roughly comparable to the size of HII regions) components \citep{JokipiiL1969, HartmanEA1997, GaenslerEA2008, SavageW2009, Azimlu, JonesEA2016}.

Despite the established importance of magnetic fields, uncertainties persist regarding their exact role in galactic evolution, primarily due to observational challenges in measuring field properties on smaller scales. This work aims to address these uncertainties by leveraging pulsar observations to refine our understanding of the Milky Way’s magneto-ionic medium, focusing on the length scale of the small-scale magnetic fields and thermal electron densities.

Pulsars serve as ideal probes of the Galaxy's magneto-ionic medium  \citep{LorimerK2012} since they provide dispersion and Faraday rotation measures, $\DM$ and $\RM$, which can be used to determine properties of both the thermal electron density and magnetic fields. For our analysis, we use $\DM$, $\RM$, and distance to the pulsar, $L$, from the ATNF Pulsar Catalogue \citep[][version 2.4.0]{ManchesterEA2005}. Their widespread distribution across the Milky Way offers great coverage of diverse Galactic environments \citep{RandK1989, HanEA2018}. Although pulsar-based measurements face challenges, particularly with distance estimation \citep{Jansson2012, Verbiest_2012, Moran_2023, KoljonenEA2024, OckerEA2024}, they remain crucial for mapping the magneto-ionic medium. Here, we first describe $\RM$ and $\DM$ (\Sec{sec:rmdmpul}) and then the method to determine the average magnetic field strength along the pulsar sightline from $\RM$ and $\DM$ (\Sec{sec:mag}).

\subsection{$\RM$ and $\DM$ from pulsars} \label{sec:rmdmpul}

The dispersion measure ($\DM$) of a pulsar is derived from the broadening of the observed pulse across a finite bandwidth. Physically, it represents the integral of the thermal electron density, $\ne$, along the line of sight from the pulsar to the observer \citep{hewish}:

\begin{equation} \label{eq:DM}
\frac{\DM}{1~\text{pc cm}^{-3}} = \int_{L} \frac{\ne}{1~\text{cm}^{-3}} \, \frac{\dd l}{1~\text{pc}},
\end{equation}
where $L$ is the total path length or equivalently the distance to the pulsar.

The rotation measure ($\RM$) quantifies the change in the polarisation angle, $\Delta \theta$, as light passes through the ISM:
\begin{equation}
\frac{\Delta \theta}{1~\text{rad}} = \frac{\RM}{1~\text{rad m}^{-2}} \, \frac{\lambda^{2}}{1~\text{m}^{2}},
\end{equation}
where $\lambda$ is the wavelength and $\RM$ is expressed as:
\begin{equation} \label{eq:RM}
\frac{\RM}{1~\text{rad m}^{-2}} = 0.812 \int_{L} \frac{\ne}{1~\text{cm}^{-3}} \, \frac{\Bp}{1~\mu\text{G}} \, \frac{\dd l}{1~\text{pc}},
\end{equation}
where $\Bp$ is the component of the magnetic field parallel to the line of sight. By combining $\RM$ and $\DM$, we can study the properties of the ISM and map magnetic fields in the Milky Way \citep[e.g.,][]{Haverkorn2015, Han2017, MaEA2020}.

\subsection{Estimating average magnetic fields from $\RM$ and $\DM$} \label{sec:mag}
By using both $\RM$ and $\DM$ from pulsars observations, we can estimate the properties of magnetic fields within the Milky Way \citep[][]{Smith1968, Manchester1972, Manchester1974, LyneS1989, RandK1989, HanMQ1999, IndraniD1999, MitraEA2003, HanEA2006, HanEA2018, SobeyEA2019, Seta2021, LeeEA2024}\footnote{Fast Radio Bursts (FRBs) can similarly be used to estimate extragalactic magnetic fields using their $\RM$ and $\DM$ values \citep{AkahoriRG2016, RaviEA2016, HacksteinEA2019, ProchaskaEA2019}.}.

The average magnetic field along the line of sight, $\Bpa$, can be estimated as
\begin{equation}\label{bp average}
\frac{\Bpa}{1~\mu\text{G}} = \frac{{\displaystyle \int_{L}} \dfrac{\ne}{1~\text{cm}^{-3}} \, \dfrac{\Bp}{1~\mu\text{G}} \, \dfrac{\dd l}{1~\text{pc}}}{{\displaystyle \int_{L}} \dfrac{\ne}{1~\text{cm}^{-3}} \, \dfrac{\dd l}{1~\text{pc}}} 
= 1.232\,\frac{\RM / (1~\text{rad m}^{-2})}{\DM / (1~\text{pc cm}^{-3})}
\end{equation}
\rev{}
This equation assumes that the thermal electron density ($\ne$) and the magnetic field ($\Bp$) are uncorrelated. If a correlation \rev{(or anti-correlation)} exists, it could lead to an overestimation \rev{(or underestimation)} of $\Bpa$ \citep{BeckEA2003}. \revst{However, o}\rev{O}ver large path lengths (on the order of kiloparsecs), this assumption has been shown to hold true \citep{Seta2021}. \rev{The effect of such a correlation (or anti-correlation) on smaller scales might exist and can be included as a correction factor \citep{BeckEA2003, ShukurovS2021} but given that most pulsars are located at such large distances, we expect the effect to be negligible \citep[consistent with][]{Seta2021}}.

Estimating the magnetic field strength along the line of sight requires knowledge of both the path length (i.e., the distance to the pulsar) and the scales over which the magnetic field varies. However, pulsar distances are often estimated indirectly via $\DM$ values, which rely on models of the Galaxy’s thermal electron density distribution. It is also often very hard to estimate the magnetic field length scale, especially for the small-scale magnetic field, which this work aims to determine.

The primary models used to estimate pulsar distances are $\mathrm{NE2001}$ \citep{NE2001} and $\mathrm{YMW16}$ \citep{YMW2013}. Here, these $\DM$-based distance estimates are referred to as $\distdm$. The ATNF pulsar catalogue \citep{ManchesterEA2005} primarily uses the $\mathrm{YMW16}$ model \citep{YMW2013}  for $\nea$, which, despite being derived from $\DM$ measurements, carries inherent uncertainties \citep{Verbiest_2012, Moran_2023}. This creates a circularity issue; the electron density model is constructed using $\DM$ values, which are then used to estimate pulsar distances. To mitigate this, the catalogue also includes independently determined distances for about 250 pulsars based on parallax measurements \citep[][version 2.4.0: \href{https://www.atnf.csiro.au/research/pulsar/psrcat}{https://www.atnf.csiro.au/research/pulsar/psrcat}]{ManchesterEA2005}. These independently determined distances are referred to as $\dista$.

Our approach extends further and diverges from previous studies by simultaneously analysing small- and large-scale thermal electron densities and magnetic fields. Our goal is to ascertain not only their strengths but also the length scale of their small-scale components, $\lb$ and $\lne$. We adopt and modify a method based on the approach outlined in Sec.~IV.2. of \cite{Ruzmaikin1988}.

The rest of the paper is organised as follows. In \Sec{sec:methods}, we describe our methodology and our results are presented in \Sec{sec:results}. In \Sec{sec:dis}, we discuss our results, their implications, and the assumptions of the study. Finally, we summarise our work and conclude in \Sec{sec:con}.

\section{Methodology} \label{sec:methods}

\subsection{Analysis of pulsar $\RM$s}\label{mainmethod1}

As briefly discussed before, the total magnetic field in galaxies, $\B_{\rm tot}$, can be divided into two components \citep[see][for a method to separate the two components from $\RM$ observations]{Seta2024}: the large-scale component, which is coherent over kiloparsec scales ($\Bm$) and the small-scale, random component, which varies over much smaller scales ($\Bf$).  The total magnetic field in galaxies can be written as,
\begin{equation}     \label{Total_B}
    \B_{\rm tot} = \Bm + \Bf.
\end{equation}
\rev{Here, we assume that the large- and small-scale magnetic field components are statistically independent of each other. We note that the small-scale component $\Bf$ can be correlated with the large-scale field $\Bm$ because turbulence also tangles the large-scale field to generate small-scale random fluctuations \citep{Hollins2017, SetaF2020, ShukurovS2021}. In our analysis, however, we treat $\Bm$ and $\Bf$ as statistically independent as it allows a tractable separation of their contributions to $\RM$. Given the lack of analytical prescription for the $\Bm$–$\Bf$ correlation due to the tangling of the large-scale field, we neglect the contribution it may introduce.}

To analyse the connection of these components individually with the observable, $\RM$, we multiply both sides of \Eq{Total_B} by the thermal electron density ($\ne$) and then take the projection along the line of sight as,
\begin{align}
    \ne \B_{{\rm tot}, \parallel} = \ne \Bp + \ne b_\parallel. \label{neb}
\end{align}

To compute $\RM$, we integrate both sides of \Eq{neb} over the path length, $L$ (here, equal to the distance to pulsar) as,
\begin{align}
    \int_L \ne \B_{{\rm tot}, \parallel} \dd l = \int_L \ne \Bp \dd l + \int_L \ne b_\parallel \dd l,
\end{align}
where 	$\Bp$ is the parallel (line-of-sight) component of the large-scale field and $b_\parallel$ is the parallel component of the small-scale field,
which gives,
\begin{equation}  \label{Rm total}
    \RM = \RM_{\Bm} + \RM_{\Bf},
\end{equation}
where $\RM$ represents the observed $\RM$ associated with the total magnetic field and $\RM_{\Bm}$ and $\RM_{\Bf}$ represents that for each component.

Adopting a simple model from Sec.~IV.2. of \citet{Ruzmaikin1988}, the contribution of the large-scale magnetic field for a single point source like a pulsar is expressed as,
\begin{align}
\RM_{\Bm}&= 0.812\,B\,\langle \ne \rangle\,L\,\cos \left(\frac{\B \cdot \vec{L}}{B L} \right), \\ 
&=\co\,L\left[\cos(\gb_0)\cos(\gb)\cos(\gl-\gl_0)+\sin(\gb_0)\sin(\gb)\right],
\label{RM_l}
\end{align}
where $\nea$ is the mean electron density along the line of sight, $L$ is the distance to the pulsar, $\vec{L}$ is the vector joining the location of the pulsar to us. The term $\co$ is defined as $0.812\,\nea\,B$. The galactic coordinates, $\gl$ and $\gb$, represent the pulsar’s position in the Galaxy, while $\gl_0$ and $\gb_0$ are coordinate reference parameters for the large-scale magnetic field model. \rev{We note that this is an unidirectional model with direction $(\gl_{0}, \gb_{0})$ and thus we primarily focus on the magnitude of the estimated large-scale field, $|B|$, for our results.} \revb{We also know that there is at least one large-scale magnetic field reversal near the Sagittarius spiral arm in the first Galactic quadrant \citep{SunEA2008, VanEckEA2011, Jansson2012, OrdogEA2017, HanEA2018} and the possible effects of this and potentially other reversals are discussed in \Sec{3.5} and \Sec{4.4}.}

\rev{We further note that, after adopting a model for the large-scale magnetic field, a perturbation-type analysis assuming the effect of the small-scale field as a weak perturbation to the large-scale field is not valid, since the observed strengths of the small-scale magnetic field, $|\boldsymbol{b}|$, are typically comparable to or even exceed those of the large-scale field, $|\boldsymbol{B}|$ \citep{Haverkorn2015, Beck2016}.}

To proceed with the statistical analysis, we first examine the mean of $\RM$, denoted $\RMa$. The mean of the total field can be expressed as the sum of the large- and small-scale contributions, 
\begin{equation}\label{Rm average}
    \RMa = \RM_{\Bm} + \langle \RM_{\Bf} \rangle,
\end{equation}
where we have used that $\langle \RM_{\Bm} \rangle \approx \RM_{\Bm}$ (see \Eq{RM_l}).

Given that the length-scale of the small-scale component is significantly smaller than the large-scale component and the distance to the pulsar (of the order of $\kpc$) and assuming uncorrelated thermal electron density and magnetic fields, we can simplify \Eq{Rm average} to
\begin{equation}\label{Rm 0}
    \langle \RM_{\Bf} \rangle \approx 0~\text{and}~\RMa \approx \RM_{\Bm}.
\end{equation} 
This implies that the total average $\RM$ is equal to the average $\RM$ of the large-scale field, as the small contribution averages out to zero. 

We continue our statistical analysis by determining the variance of $\RM$ or how much $\RM$ fluctuates. This is given by
\begin{align}
    \varsigma^{2}(\RM) = \varsigma^{2}(\RM_{\Bm}) + \varsigma^{2}(\RM_{\Bf}).
\label{rm_sigma}
\end{align}
Since most variations in $\RM$ along the line of sight are due to the small-scale, random magnetic field fluctuations,
\begin{align}
\varsigma^{2}(\RM_{\Bm}) \approx 0 ~\text{and}~ \varsigma^{2}(\RM) \approx \varsigma^{2}(\RM_{\Bf}).
\label{rm_sigma2}
\end{align}
This implies that the $\RM$ fluctuations are primarily determined by the small-scale magnetic field fluctuations.

Now, we introduce the quantity,
\begin{equation}
    \mathcal{P}(l) = \ne(l)~(\B(l) \cdot \hat{L}),
    \label{ne}
\end{equation}
where $\hat{L}$ represents the unit vector along the line of sight and $l$ denotes distance along it from the pulsar. Using $\mathcal{P}(l)$, the mean and variance of $\RM$ can be written as
\begin{equation}\label{eq:in}
    \RMa = 0.812\,\int_{L}\,\mathcal{P}(l)\, \dd l
\end{equation}
and
\begin{equation}
   \varsigma^{2}(\RM) = (0.812)^2\,\int_{L} \dd l_1\int_{L} \dd l_2 \left[ \overline{\mathcal{P}(l_2)\mathcal{P}(l_2)} - \bar{\mathcal{P}}(l_1)\bar{\mathcal{P}}(l_2) \right]. 
    \label{var}
\end{equation}

\rev{In \Eq{var}, the overbars denote an ensemble or spatial average over realisations of the random magnetic field or electron density along a line of sight, while the angular brackets $\langle \cdot \rangle$ (e.g.~in \Eq{Rm 0}) denote a statistical or expected value. In practice, both averaging procedures are related, but we use overbars to indicate a local or sample-based mean and angular brackets to indicate a theoretical or global expectation value.}

Assuming that the small-scale field is homogeneous, the quantity within the square bracket in \Eq{var} is a two-point correlation function $\mathcal{C} (l_1, l_2)$. This is a crucial assumption that allows the correlation function to depend only on the distance between two points \citep[this is also directly related to structure functions, described in][]{SetaEA2023},
\begin{equation} \label{sep}
    s = |l_1-l_2|.
\end{equation}

After some algebra, \Eq{var} simplifies to,
\begin{equation}
    \varsigma^{2} (\RM) =2\,\co^{2}\,\int_{L} (L-s)\, \mathcal{C}(s)\, \dd s.
\end{equation}

\rev{This analysis assumes the presence of a characteristic scale of the small-scale field, $\ell_{b}$ (the typical size over which the small-scale magnetic field remains correlated). When $s \ll \ell_{b}$, the values of $\mathcal{P}(s_1)$ and $\mathcal{P}(s_2)$ are strongly correlated, resulting in larger values of $\mathcal{C}(s)$. Conversely, for $s \gg \ell_{b}$, $\mathcal{P}(s_1)$ and $\mathcal{P}(s_2)$ become effectively uncorrelated, so that $\mathcal{C}(s)\to 0$. In physical terms, if two points along the line of sight are separated by a distance smaller than $\ell_{b}$, the magnetic fields at those points are still related. If the distance between them is much larger than $\ell_{b}$, the magnetic fields at those points no longer exhibit measurable correlation. We emphasise that here we are referring to the absence of correlation, which does not necessarily imply full statistical independence. }

For a simple model, we assume correlation $\mathcal{C}(s)$ is represented by an exponential function,
\begin{equation}
    \mathcal{C}(s) = \Co \exp(-s/\lb),
    \label{expo}
\end{equation}
where $\Co$ is a dimensionless constant. This correlation decays over a characteristic scale length, $\lb$, which represents the typical size of magnetic field structures or the scale of small-scale magnetic fields. By adopting this exponential form, we can express the variance in $\RM$ (\Eq{var}) that arises from small-scale magnetic fields (see \App{app1a} for the full derivation) as,
\revst{This can be further simplified into}
\begin{align}\label{rm3}
    \varsigma ({\RM})&={2}^{1/2}\,\co\,{C_0}^{1/2}\,\lb \left[L/\lb - 1 + \exp(-L/\lb) \right]^{1/2}.
\end{align}

Now, since $\langle \RM \rangle \approx \RM_{\Bm}$ and $\sigma (\RM) \approx \sigma (\RM_{\Bf})$, using \Eq{RM_l} and \Eq{rm3}, for each pulsar, we model the total observed $\RM$ as
\begin{align}\label{rm4}
\RM & = \RM_{\Bm} + \RM_{\Bf} \nonumber  \\ 
 & = \co L [\cos(\gb_0) \cos(\gb) \cos(\gl - \gl_0) + \nonumber \\ & \sin(\gb_0) \sin(\gb)] + \left[L/\lb - 1 + \exp(-L/\lb) \right]^{1/2} \epsilon_{\RM},
\end{align}
where $\epsilon_{\RM}$ is a Gaussian random number picked from a distribution with mean zero and standard deviation $= {2}^{1/2}\,\co\,{C_0}^{1/2}\,\lb$ to capture the contribution from the small-scale magnetic fields to the total observed $\RM$. 

The observed value of $\RM$ is now parameterised using four values ($\co$, $\gl_0$, $\gb_0$ and $\lb$) of the modelled small- and large-scale magnetic fields. Note that with each $\RM$ value, the location ($\gl$ and $\gb$) of and distance ($L$) to each pulsar are known in \Eq{rm4}. Given a sample with $\ndata$ pulsars, to ensure $\RMa \approx \RM_{\Bm}$ and $\RMa_{\Bf} \approx 0$, we minimise the combined contribution of the sum of all $\epsilon_{\RM}$s added in Gaussian quadrature to the observed $\RM$, $\mathcal{S}_{\RM}$, to determine the parameters, $\co, \gl_0, \gb_0$, and $\lb$, where $\mathcal{S}_{\RM}$ is given as,
\begin{align} \label{regressionmethod}
\mathcal{S}_{\RM}\,(\co, \gl_0, \gb_0,\lb) & = \frac{1}{\ndata} \sum_{i=1}^{\ndata} (\epsilon_\RM)_{i}^{2} \nonumber \\ 
& = \frac{1}{\ndata} \sum_{i=1}^{\ndata} \left(\frac{\RM_i - \RM_{\Bm,i}}{[L_i/\lb - 1 + \exp(-L_i/\lb)]^{1/2}}\right)^2,
\end{align}
where $\RM_i$ and $L_i$ are the observed $\RM$ of and distance to the $i^{\text{th}}$ pulsar in the sample with a total of $\ndata$ pulsars and $\RM_{\Bm, i}$ is the contribution  of the large-scale field for the $i^{\text{th}}$ pulsar computed using \Eq{RM_l}.

\subsection{Analysis of pulsar $\DM$s}\label{mainmethod2}

\revst{Like magnetic fields, the thermal electron density can also be divided into large- (more diffuse) and small- (more clumpy, in and around HII regions) scale components as,}

\rev{Similar to the magnetic field decomposition, the thermal electron density can be separated into large- and small-scale components. The large-scale part, $\langle \ne \rangle$, represents variations in the diffuse component varying over possibly larger $\kpc$-sized ISM regions, while the small-scale fluctuations, $\delta \ne$, capture clumpy structures on scales of $10$ -- $100$s of $\,\pc$ and probably also localised enhancements near HII regions \citep[size of HII regions varies over a wide range,  $10$ -- $200\,\pc$, see][]{Azimlu}. This decomposition can be written as,}
\begin{equation}
\ne = \langle \ne \rangle + \delta\ne.
\label{ne_total}
\end{equation}
To compute the total $\DM$ we integrate both sides of \Eq{ne_total} over the distance $L$ to the pulsar as
\begin{align}\label{Dm total}
\int_L \ne\, \dd l &= \int_L \langle \ne \rangle\, \dd l + \int_L \delta \ne\, \dd l, \\
\DM &= \DM_{\nea}+ \DM_{\delta \ne}.
\end{align}
Here, $\DM_{\nea}$ represents the contribution from the large-scale, diffuse component of the thermal electron density, while $\DM_{\delta \ne}$ captures the contribution from small-scale fluctuations. 

Like we did with the $\RM$s, our statistical analysis of $\DM$ begins with their mean,
\begin{equation}
    {\DMa = \langle\DM_{\nea}\rangle + \langle \DM_{\delta \ne}\rangle}
\end{equation}
\rev{where $\langle\DM_{\nea}\rangle$ represents the mean contribution from large-scale thermal electron density structures and $\langle \DM_{\delta \ne}\rangle$ represents the fluctuating contribution from small-scale thermal electron density. Again, we assume that $\DMa$ is primarily determined by $\langle \DM_{\delta \ne}\rangle$, with $\langle \DM_{\delta \ne}\rangle$ contributions being negligible in comparison \citep[see][for further discussion on this assumption]{JokipiiL1969, NelemansEA1997}}. Thus,
\begin{align}
\langle \DM_{\delta \ne}\rangle \approx 0~\text{and}~\DMa \approx \langle\DM_{\nea}\rangle.
\end{align}
Furthermore, 
\begin{align}
\langle\DM_{\nea}\rangle = \nea\,L.
\label{dmmean}
\end{align}

Then computing the variance gives
\begin{align}
\varsigma^{2}(\DM) = \varsigma^{2}(\DM_{\nea}) + \varsigma^{2}(\DM_{\delta \ne}).
\label{Dm_sigma}
\end{align}
Assuming that the $\varsigma^{2}(\DM)$ is primarily due to the small-scale thermal electron density, we get
\begin{align}
\varsigma^{2}(\DM_{\nea}) \approx 0~\text{and}~\varsigma^{2}(\DM) \approx \varsigma^{2}(\DM_{\delta \ne}).
\end{align}
The term $\varsigma^{2}(\DM) \approx \varsigma^{2}(\DM_{\delta \ne})$ is now modelled below.

We now introduce,
\begin{equation}\label{eq:in2}
\mathcal{Q}(l) = \ne (l).
\end{equation}
To model the total observed $\DM$ following the same approach as used for the $\RM$ analysis in \Sec{mainmethod1}, the variance of $\DM$ can be expressed as,
\begin{equation}\label{var2}
    \varsigma^{2} (\DM) = \int_{L} \dd l_1 \int_{L} \dd l_2 \left[\overline{\mathcal{Q}(l_1)\mathcal{Q}(l_2)} - \bar{\mathcal{Q}}(l_1)\bar{\mathcal{Q}}(l_2)\right].
\end{equation}
This can be further reduced in terms of the correlation function of the small-scale thermal electron density as,
\begin{equation}
    \varsigma^{2} (\DM) =2\,\nea^{2}\int_{L} (L-s)\, \mathcal{C}(s)\, \dd s.
\end{equation}
where $\mathcal{C}(s)$ represents the correlation function of the small-scale fluctuations. Assuming the familiar exponential decay for the correlation function,
\begin{equation}
    \mathcal{C}(s) = \Co \exp(-s/\lne),
    \label{expo2}
\end{equation}
we obtain,
\begin{equation}
    \varsigma^{2} ({\DM}) = 2\,\nea^{2}\left[L\lne + \lne^{2}\left(-1 +\exp(-L/\lne)\right) \right], 
\end{equation}
which gives,
\begin{align}
    \varsigma ({\DM})&=2^{1/2}\,{C_0}^{1/2}\,\nea\,\lne \left[L/\lne - 1 + \exp(-L/\lne) \right]^{1/2}.
    \label{dmfluc}
\end{align}

Thus, using equation \Eq{dmmean} and equation \Eq{dmfluc}, the observed total $\DM$ is written as
\begin{align}\label{dm4}
\DM &= \DM_{\nea}+ \varsigma (\DM) \nonumber \\ 
 & = \nea\, L + \left[L/\lne - 1 + \exp(-L/\lne) \right]^{1/2} \eta_{\DM}
\end{align}
where $\eta_{\DM}$ is a Gaussian random number picked from a distribution with mean zero and standard deviation $={2}^{1/2}\,{C_0}^{1/2}\,\nea\,\lne$.

The observed $\DM$ for a pulsar at a distance $L$ is parameterised using two values, $\nea$ and $\lne$, which are properties of the thermal electron density model. Given a sample of $\ndata$ pulsars, to ensure that we minimise the combined contribution of the sum of all  $\eta_{\DM}$ added in Gaussian quadrature to the observed $\DM$, $\mathcal{S}_{\DM}$, to determine the parameters, $\nea$ and $\lne$, where $\mathcal{S}_{\DM}$ is given as,
\begin{align} \label{regressionmethoddm}
\mathcal{S}_{\DM}\,(\nea,\lne) & = \frac{1}{\ndata} \sum_{i=1}^{\ndata} (\eta_\DM)_{i}^{2} \nonumber \\ 
& = \frac{1}{\ndata} \sum_{i=1}^{\ndata} \left(\frac{\DM_i - \DM_{\nea,i}}{[L_i/\lne - 1 + \exp(-L_i/\lne)]^{1/2}}\right)^2,
\end{align}
where $\DM_i$ and $L_i$ are the observed $\DM$ of and distance to the $i^{\text{th}}$ pulsar in the sample, respectively. $\DM_{\nea,i}$ represents the large-scale thermal electron density contribution to the $\DM$ for the $i^{\text{th}}$ pulsar, $\DM_i$, which is calculated using \Eq{dmmean}.

\subsection{Correlation functions}\label{mainmethod3}

\begin{figure}
\includegraphics[width=\columnwidth]{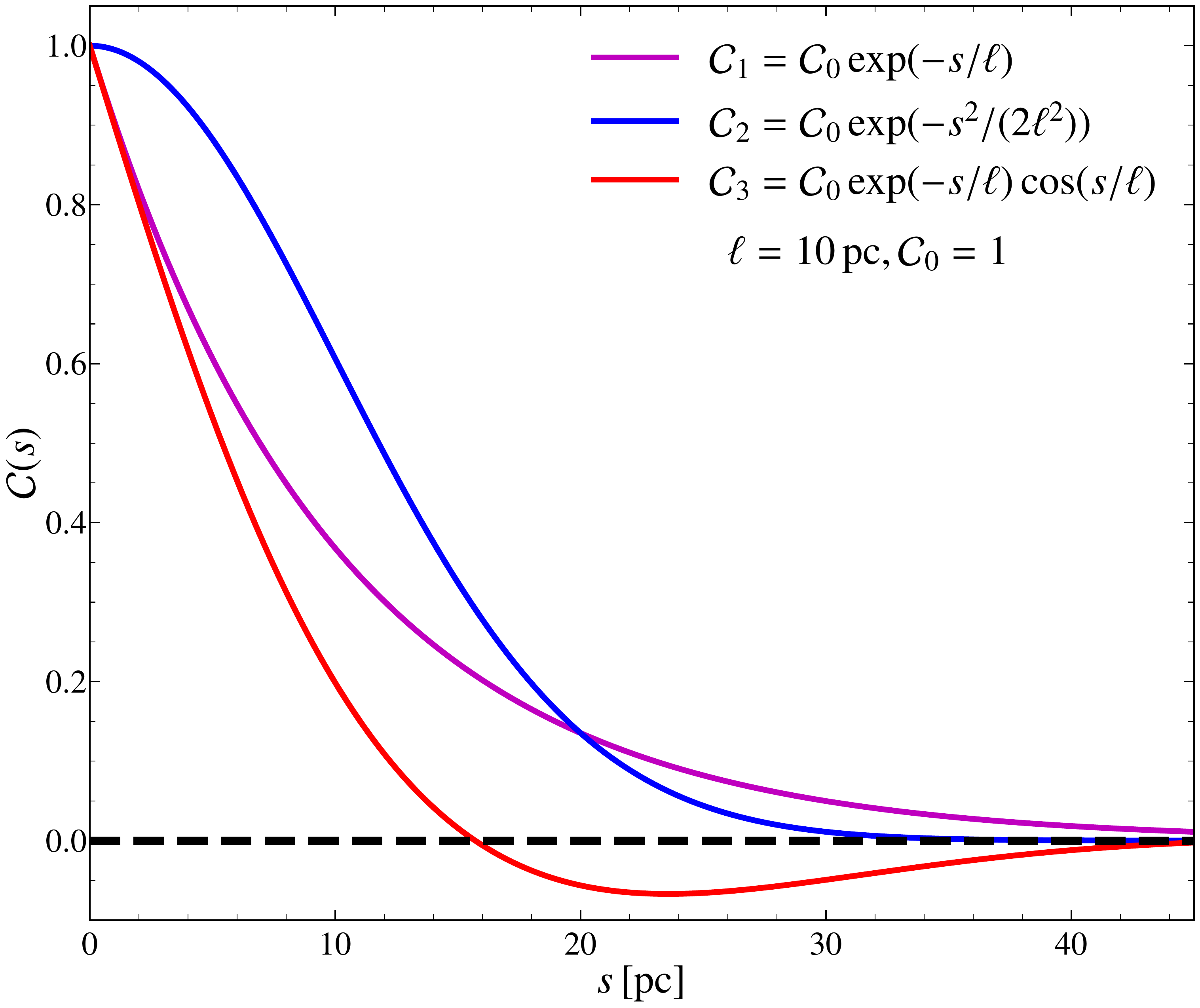}
\caption[Comparison of correlation functions]{Comparison of three different correlation functions, $\Ca$, $\Cb$, and $\Cc$ (forms in the legend), as functions of the separation distance, $s$, in $\pc$ with a characteristic scale length, $\ell = 10\,\pc$ and the dimensionless constant, $\Co=1$. All the correlation functions start from $1$ at $s=0$ and approach $0$ (dashed, black line) as $s \to \infty$. $\Cb$ goes to zero at a smaller separation than $\Ca$ and $\Cc$ allows the correlation to be negative before approaching zero. This illustrates the distinct behaviours of each correlation function at different separation distances, emphasising their functional forms, relative amplitudes, and decay rates (see \Sec{mainmethod3} for further discussion on differences).}
\label{corfunc}
\end{figure}

One of the crucial assumptions in our statistical analysis is the choice of the correlation function for the small-scale components, $b$ and $\delta \ne$. Though \citet{Ruzmaikin1988} used only an exponentially decaying function (e.g.~\Eq{expo}), motivated by magnetic field correlation functions obtained from the results of turbulent magnetohydrodynamic simulations of the ISM \citep{Hollins2017}, we use the following three correlation functions for both the $\RM$ and $\DM$ analysis,
\begin{equation}
\begin{aligned}
    \Ca (s) &= \Co \exp(-s/\ell), \\
    \Cb (s) &= \Co \exp(-s^2/(2 \ell^2)),\\
    \Cc (s) &= \Co \exp(-s/\ell) \cos(s/\ell),
\end{aligned}
\label{c func}
\end{equation}
where $s$ denotes separation (\Eq{sep}), $\ell$ is the characteristic length scale (either $\lb $ or $\lne$), and $\Co$ is a dimensionless constant. For a typical length scale, $\ell = 10\,\pc$, \Fig{corfunc} shows how all three correlation functions depend on the separation, $s$. 

For all three cases, the correlation is high at smaller separations and goes to zero at large separations, but the differences capture possible expected variations in the ISM \citep[see][for further details]{Hollins2017}. For example, the first two functions are decaying but $\Cb$ decays faster than $\Ca$ (compare blue and red lines in \Fig{corfunc}). 
\rev{Also, $\Ca$ and $\Cb$ model simple attenuation with separation, while the combination of an exponential decay with a cosine term in $\Cc$ allows the correlation to be negative and still bounce back to zero at larger distances. This anticorrelation in $\Cc$ can arise naturally from the solenoidal property of magnetic fields \citep[see Sec.~8.IV in][]{Zeldovich}.} \revb{Furthermore, for $\Cc$, the same scale is used in both the exponential and cosine terms. This scale is close to the correlation length of the random fields and, for the magnetic fields, this form of the correlation function assumes that typical field lines close within one correlation length. Such an} approach provides a potentially more flexible representation of the small-scale components, enhancing our ability to model the complexities expected in the magnetic fields and thermal electron density distributions. 

\rev{For each correlation function, the expressions used to compute the residual sums, $\mathcal{S}_{\RM}$ (\Eq{regressionmethod}) and $\mathcal{S}_{\DM}$ (\Eq{regressionmethoddm}), are derived analytically in \App{app1}. Importantly, the last term in the denominator of these expressions, which reflects the contribution of the small-scale fluctuations, is explicitly modified depending on the chosen correlation function. This ensures that $\mathcal{S}_{\RM}$ and $\mathcal{S}_{\DM}$ correctly account for the different functional forms, whether $\Ca$, $\Cb$, or $\Cc$.} Our method remains fully analytical up to this point, which motivates the choice of these three correlation functions, as each allows for a closed-form expression suitable for residual sums and subsequent numerical minimisation (described in the next subsection).

\subsection{Minimisation}\label{mainmethod4}

Assuming various possible values of $\lb$ and observational data for 27 pulsars, \citet{Ruzmaikin1988} used analytical minimisation for parameters $\gl_0$ (assuming $\gb_0=0^\circ$), $\nea$, and $B$. For our analysis, we use the $\texttt{lmfit}$ library \citep{NewvilleEA2015} for numerical minimisation (including nonlinear methods). In \App{app2}, we replicated their work on the original dataset using our method. Here, we expand the method using our larger dataset ($\ndata \sim 3000$ for $\DM$ and $\ndata \sim 1300$ for $\RM$), extending minimisation to include all parameters ($\gl_0$, $\gb_0$, $\nea$, $B$, $\lne$, and $\lb$) and for all three correlation functions given in \Sec{mainmethod3}.

However, this approach yielded an excessively low reduced chi-squared value ($\redchi \ll 1$), indicating overfitting. Adjustments with varying correlation functions (\Sec{mainmethod3}) did not improve $\redchi$ significantly, suggesting that our dataset size may not yet support the full parameter set. Thus, we fixed $\lb$ and $\lne$ at predefined scales, iterating over $\ell$ values from $0.1\,\pc$ to $1\,\kpc$ as
\begin{align*}
\ell\,[\pc] = [0.1, 1, 5, 10, 15, 20, 25, 30, 50, 100, 200, 250, 500, 1000].
\end{align*}
For each $\ell$, we applied numerical least-squares minimisation for $\gl_0$, $\gb_0$, $\nea$, and $B$, assessing fit quality with $\redchi$. Then we interpolated $\redchi$ to refine $\lb$ and $\lne$ estimates, identifying scales where $\redchi$ approaches 1, optimising $\gl_0$, $\gb_0$, $\nea$, and $B$.

To quantify uncertainties, we employed a Monte Carlo-type approach, perturbing $\RM$ and $\DM$ data up to 30,000 times based on observational errors ($\RM_{\text{ERR}}$ and $\DM_{\text{ERR}}$) and distance uncertainties \rev{(if available)}. Each perturbed dataset was fit and parameter uncertainties were estimated as the difference between the 84th and 16th percentiles.

\subsection{Data}\label{dataset}

The data for this work was sourced from the Australian National Telescope Facility (ATNF) pulsar catalogue \citep[][version 2.4.0, \href{https://www.atnf.csiro.au/research/pulsar/psrcat}{https://www.atnf.csiro.au/research/pulsar/psrcat}]{ManchesterEA2005}. We restricted the dataset to pulsars within the Milky Way, specifically those with distances $\leq 20\,\kpc$. The primary parameters of interest include:
\begin{enumerate}
\item {$\gl$:} Galactic longitude $[\degree]$
\item {$\gb$:} Galactic latitude $[\degree]$
\item {$\RM$:} Rotation measure $[\rad \,\m^{-2}]$
\item {$\DM$:} Dispersion measure $[\pc \,\cm^{-3}]$
\item {$\distdm$:} Distance to the pulsar based on the electron density model by \citet{YMW2013} [$\kpc$]
\item {$\dista$:} Independently determined distance to the pulsar [$\kpc$]
\end{enumerate}
The dataset includes 2996 entries with both $\DM$ and $\distdm$ values, while $\RM$ measurements are available only for 1290 pulsars. The subset of pulsars with independent distance estimates ($\dista$) includes 375 entries with $\DM$ values and 218 of these have $\RM$ values. It is also important to note that the parameters $\RM$, $\DM$, and $\dista$ have symmetric errors associated with them \citep[we note that $\dista$ might have asymmetric errors for some of the pulsars see,][but they are currently not included in the version 2.4.0 of the ATNF pulsar catalogue]{DellerEA2019}, while the relative errors in $\gl$ and $\gb$ computed from corresponding right ascension and declination are negligible, and $\distdm$ do not have any reported errors in the dataset.

\section{Results} \label{sec:results}

\begin{figure*}
\includegraphics[width=8.5cm, height=8.5cm]{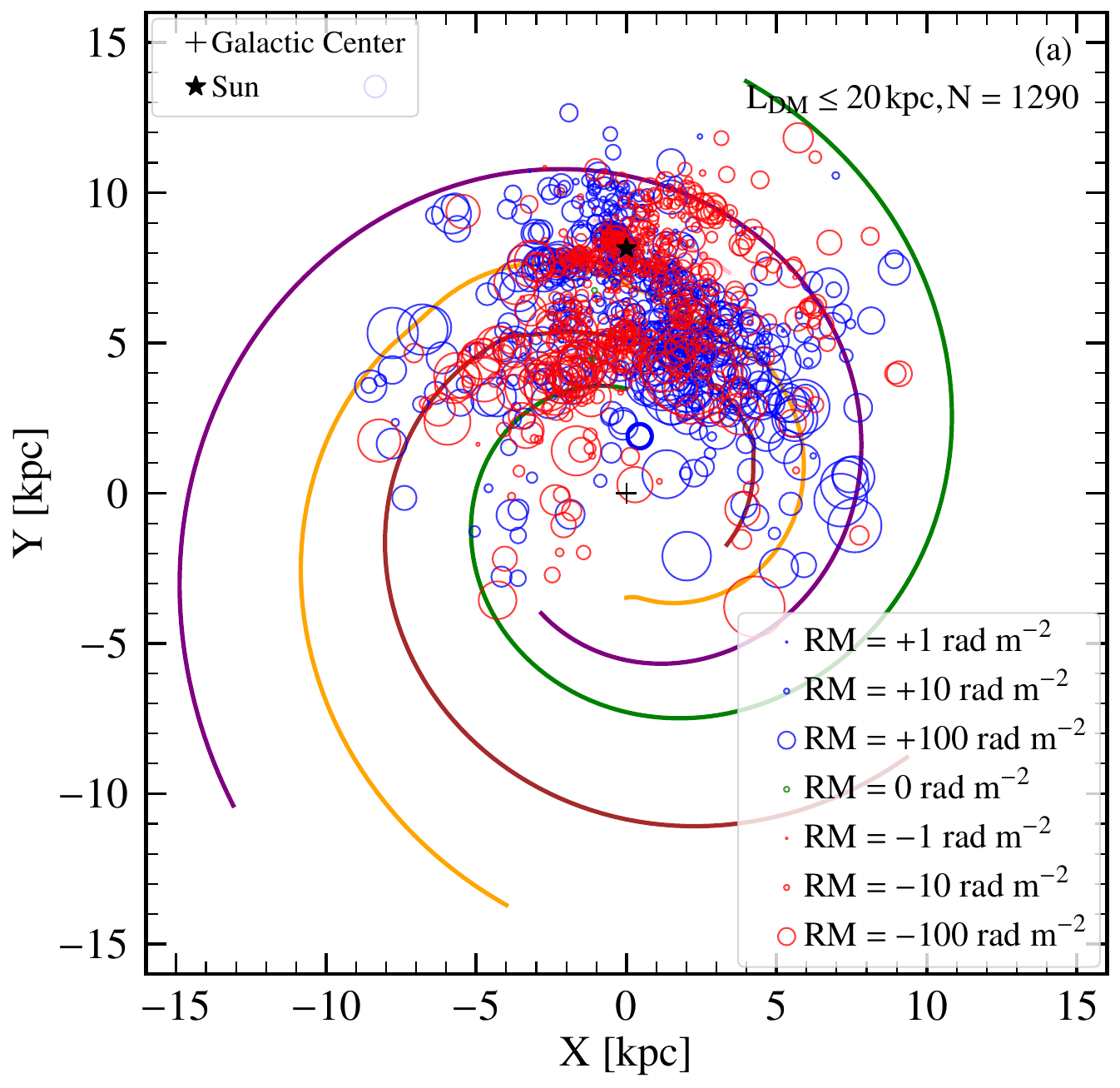}
\includegraphics[width=8.5cm, height=8.5cm]{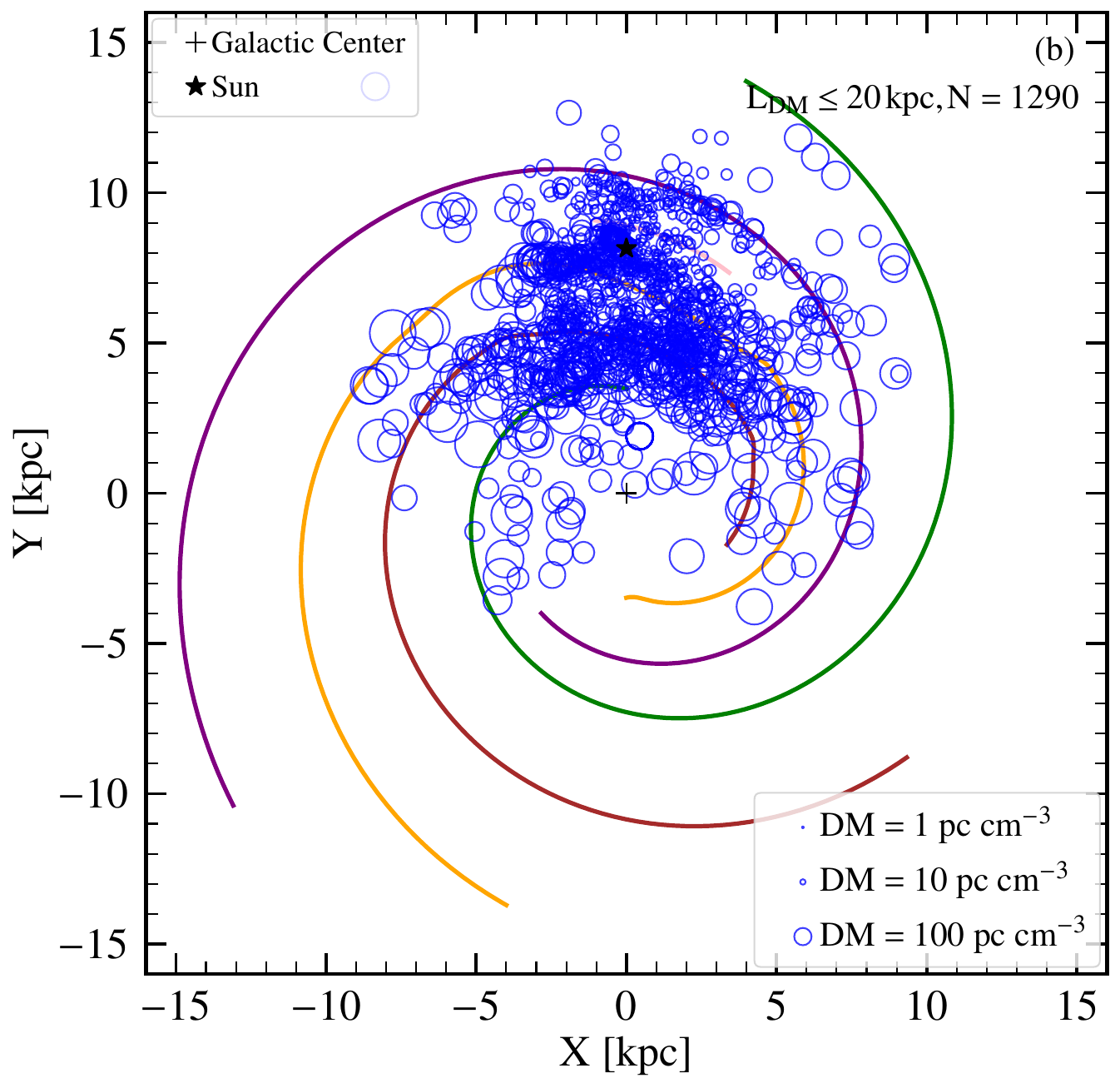}
\includegraphics[width=8.5cm, height=8.5cm]{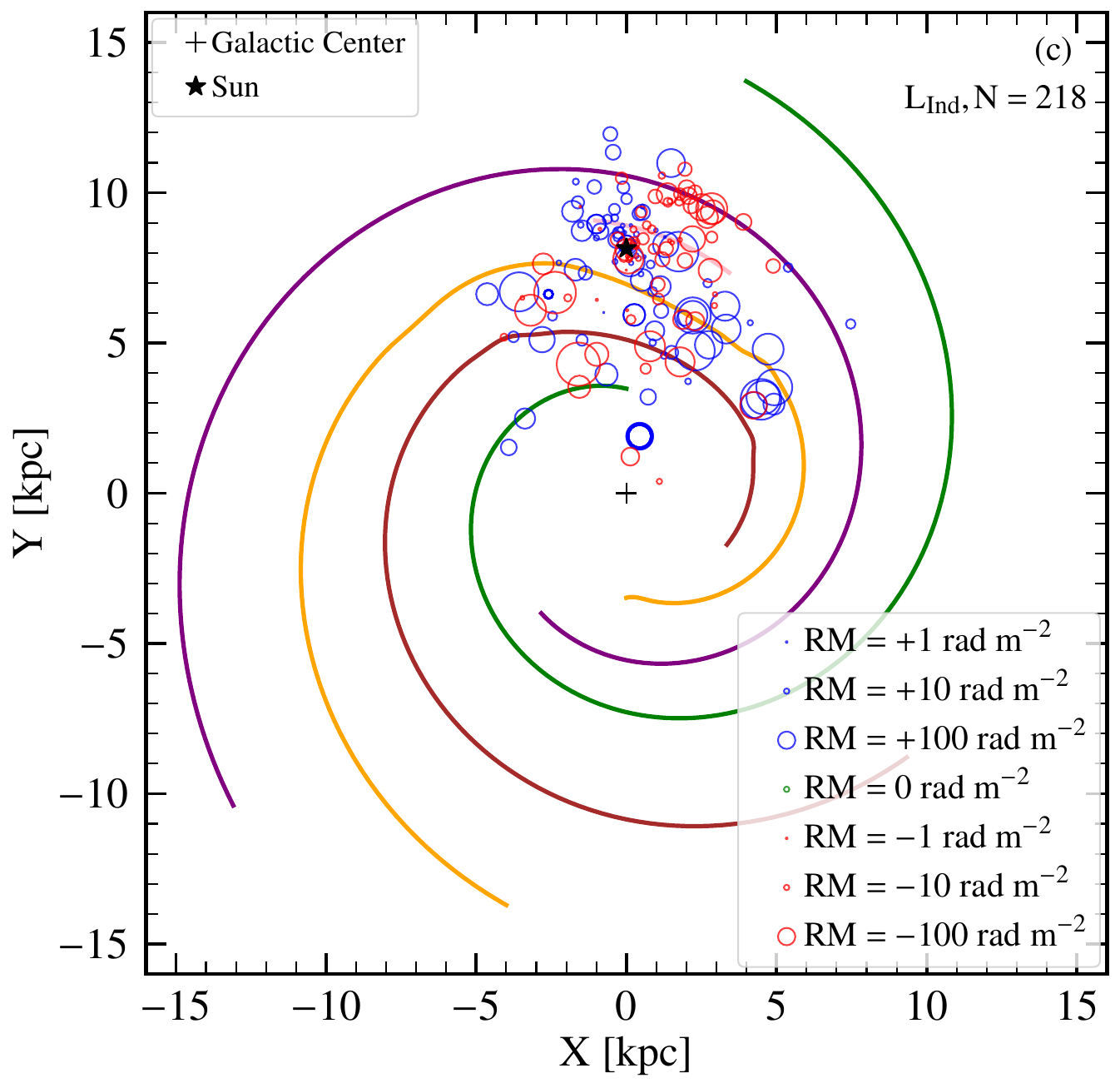}
\includegraphics[width=8.5cm, height=8.5cm]{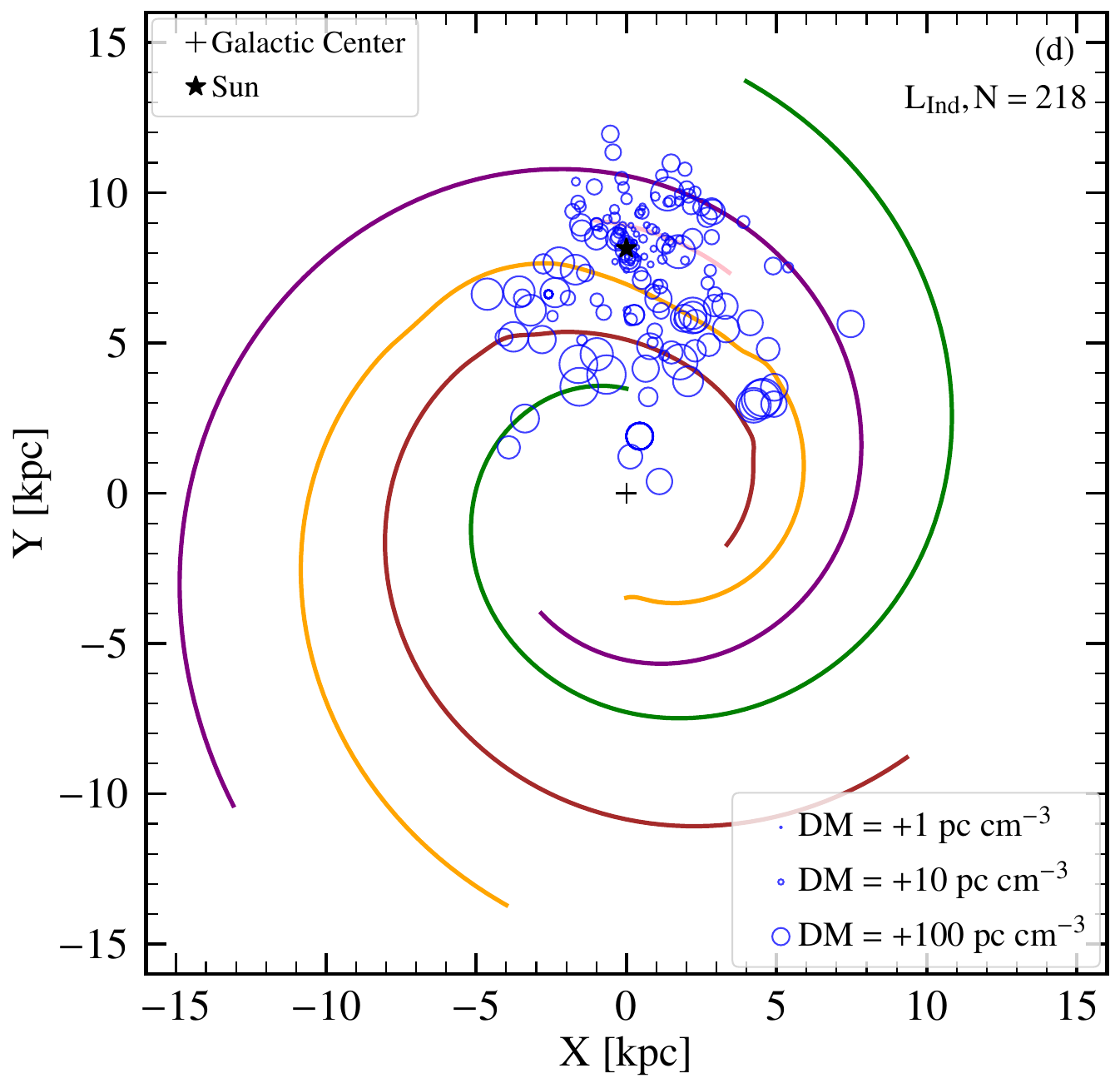}
\caption{\rev{Pulsar data, $\RM$ (a, c) and $\DM$ (b, d), for both data sets, $\distdm \le 20\,\kpc$ (a, b) and $\dista$ (c, d) with spiral arm model used in the NE2001 thermal electron density model \citep{NE2001}. As expected, most detected pulsars are concentrated around the Sun and thus the surrounding area is more sampled by the data. Our statistical analysis is, by extension, probing those regions more compared to the other parts of the Milky Way.}}
\label{spirals}
\end{figure*}

\subsection{Distribution of data}

\subsubsection{Pulsar distribution with spiral arms} \label{sec:pulspiral}
\rev{ 
\Fig{spirals} shows the distribution of pulsar data with spiral arms taken from the model used in the NE2001 thermal electron density model \citep{NE2001}. Most pulsars lie closer to the Sun and thus our statistical analysis, detailed in \Sec{sec:methods}, is more sensitive to this part of the Milky Way compared to others. We also note that some of the distances in the $\distdm$ sample can actually be incorrect and some of the pulsars can be closer than given by $\distdm$ \citep{KoljonenEA2024, OckerEA2024}. This is also the reason we perform a parallel analysis with the $\dista$ sample but there the sample size is significantly smaller (by a factor of $\approx 6$) compared to the $\distdm$ sample and the $\dista$ dataset has even sparser sampling as seen in \Fig{spirals}~(c, d). Thus, for completeness, we provide and discuss results for both samples.
}

\subsubsection{Distance distribution: $\distdm$ and $\dista$}
\begin{figure*}
    \centering
    \begin{minipage}{0.32\textwidth}
        \centering
        \includegraphics[width=\textwidth]{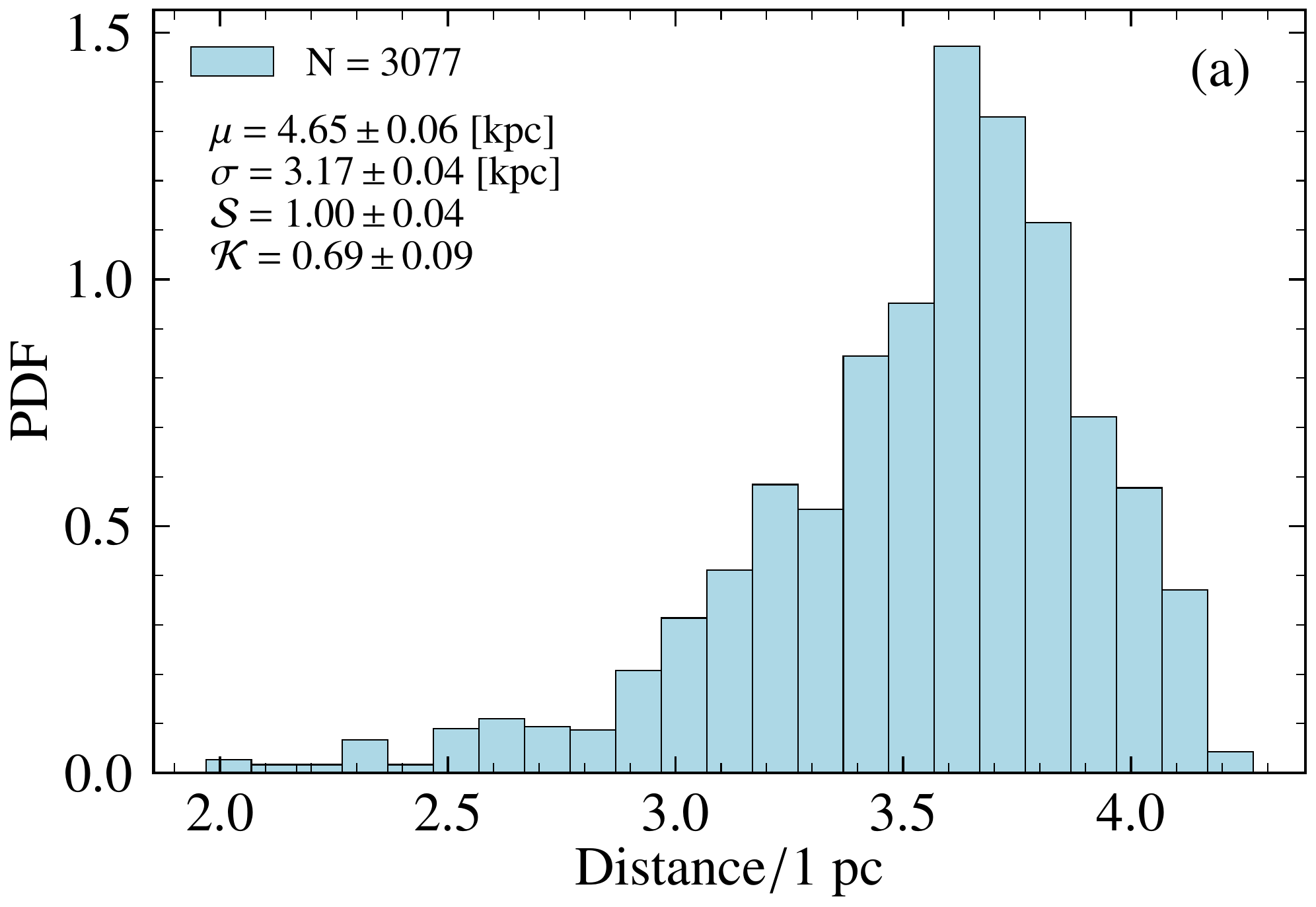}
    \end{minipage}\hfill
    \begin{minipage}{0.32\textwidth}
        \centering
        \includegraphics[width=\textwidth]{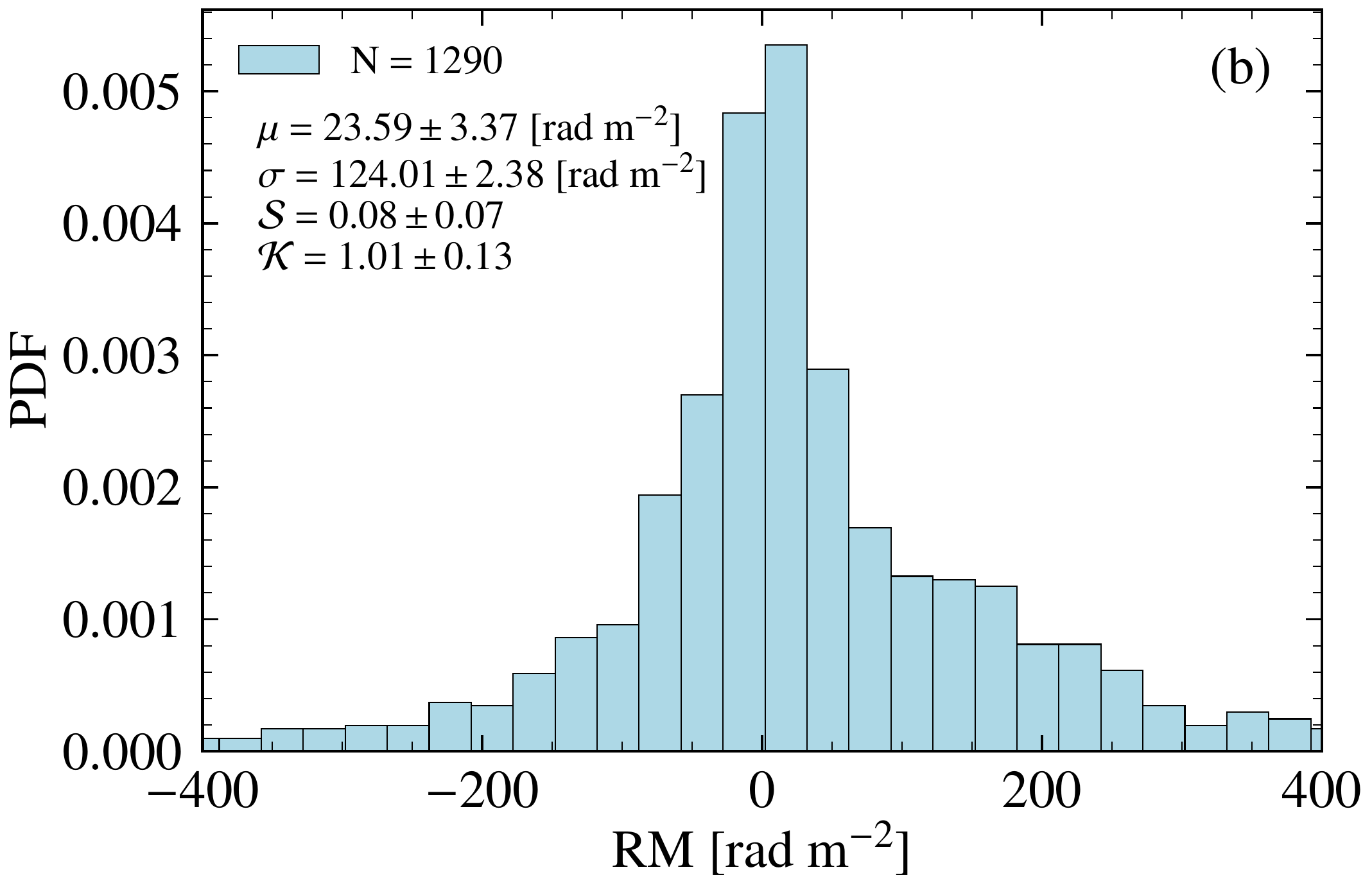}
    \end{minipage}\hfill
    \begin{minipage}{0.32\textwidth}
        \centering
        \includegraphics[width=\textwidth]{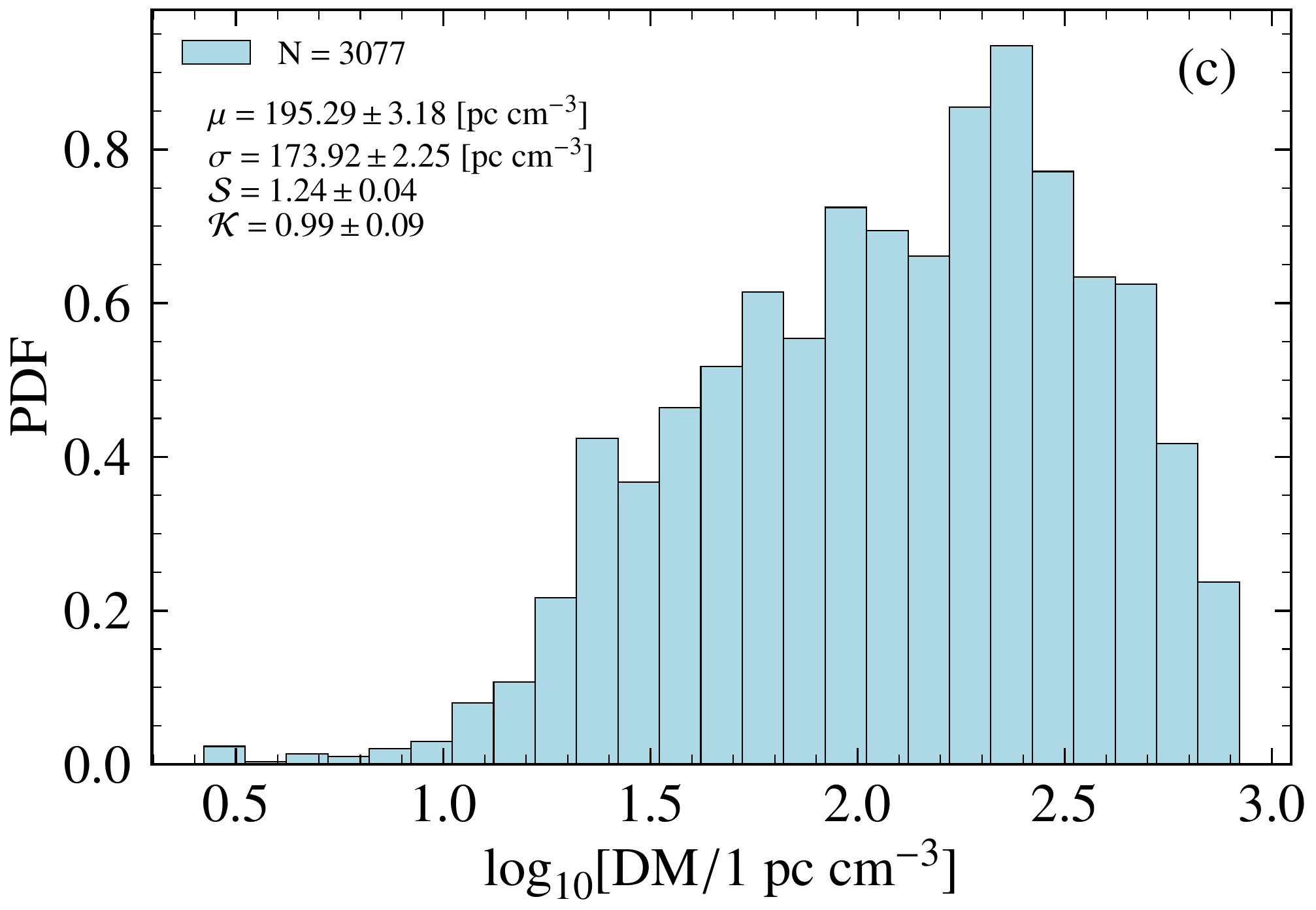}
    \end{minipage}
    \caption{Distribution of data for $\RM$, $\DM$, and $\distdm$. Figures~(a),~(b) and~(c) illustrate the histograms of $\RM$, $\DM$, and $\distdm$, respectively. All distributions observe moderate skewness except ~(b), which is more symmetric. In all figures, we also observe a $\kurt > 0$, suggesting that the data exhibits features of a non-normal distribution.}
    \label{data distdm}
\end{figure*}

\begin{figure*}
    \centering
    \begin{minipage}{0.32\textwidth}
        \centering
        \includegraphics[width=\textwidth]{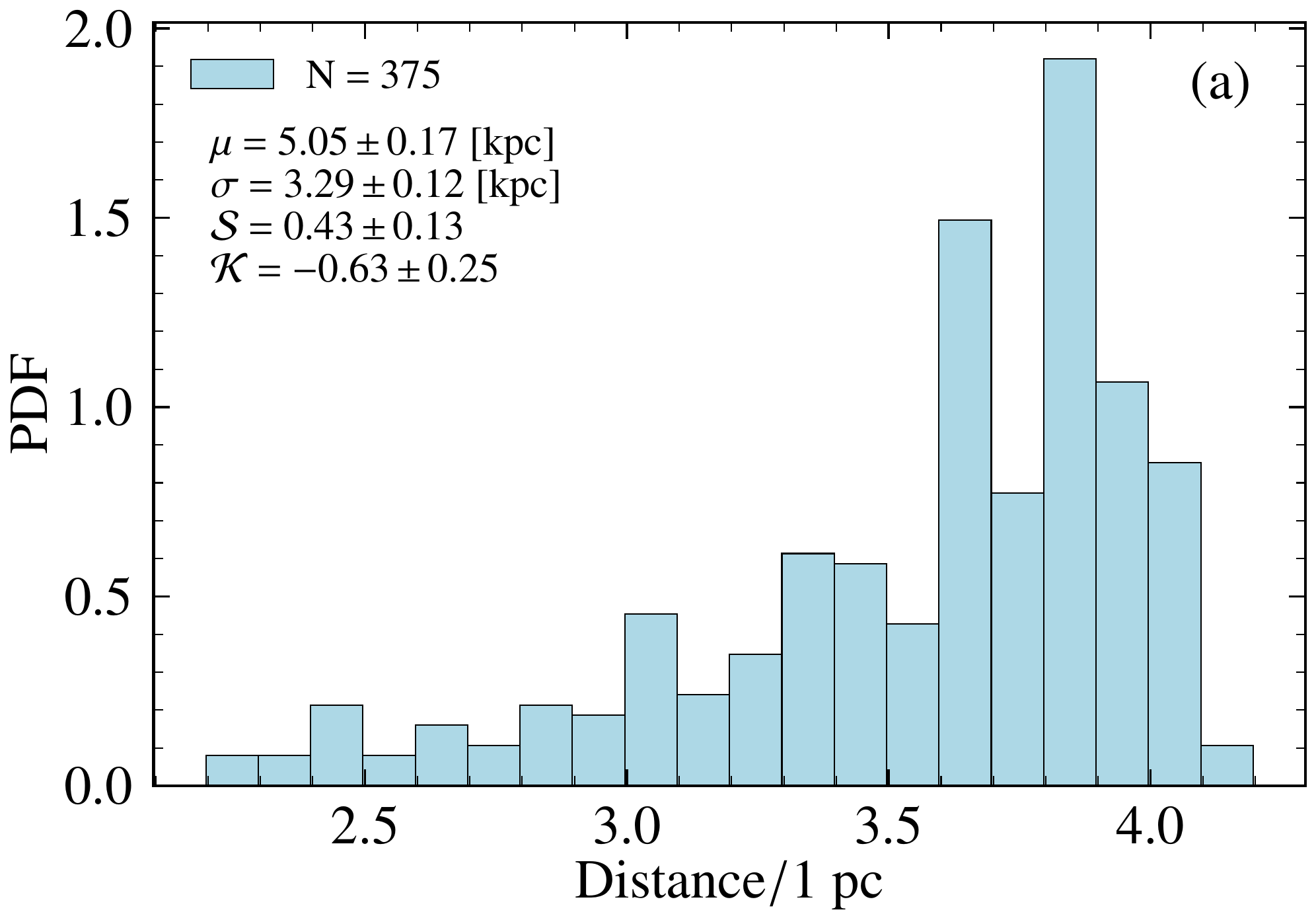}
    \end{minipage}\hfill
    \begin{minipage}{0.32\textwidth}
        \centering
        \includegraphics[width=\textwidth]{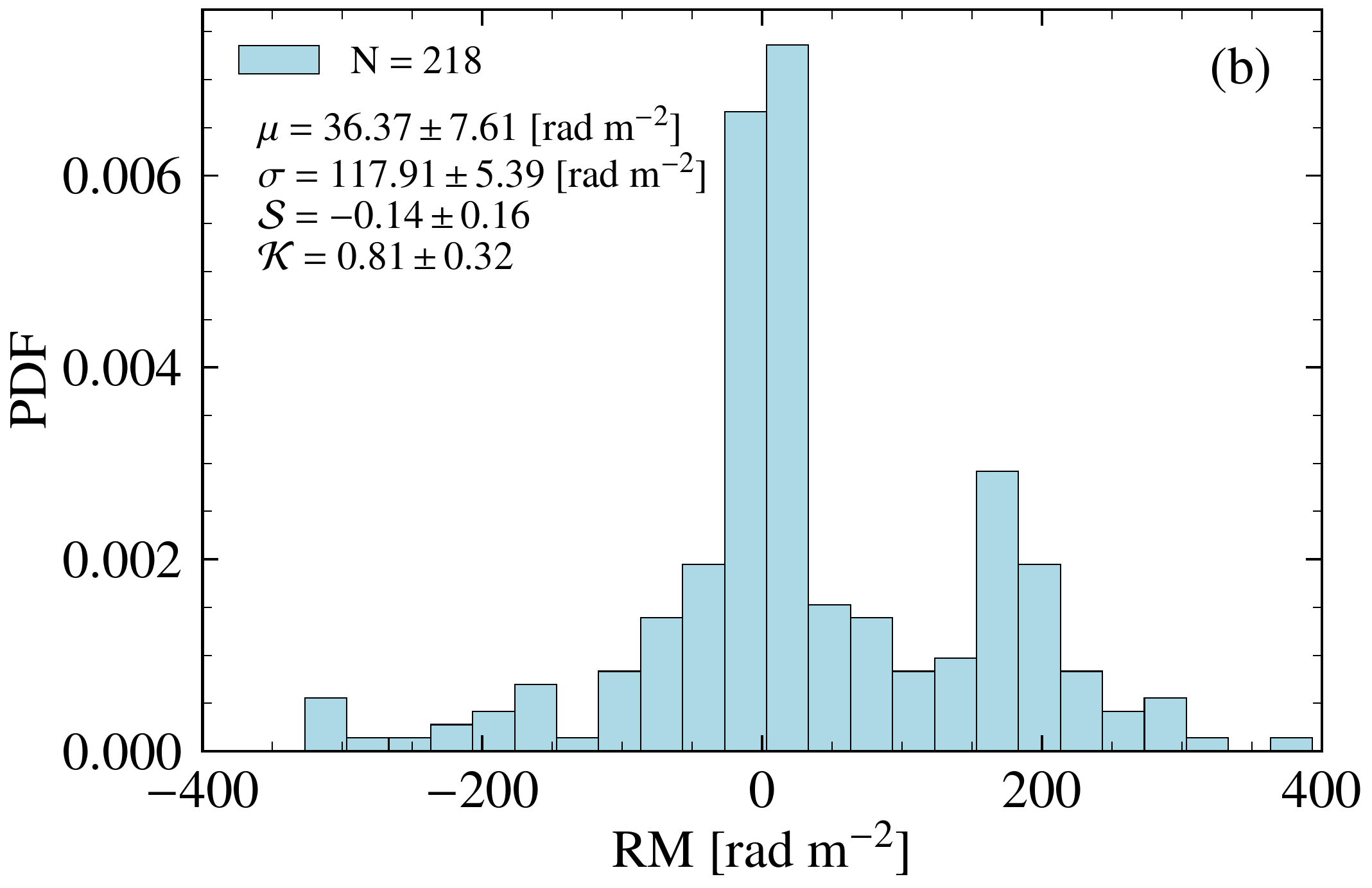}
    \end{minipage}\hfill
    \begin{minipage}{0.32\textwidth}
        \centering
        \includegraphics[width=\textwidth]{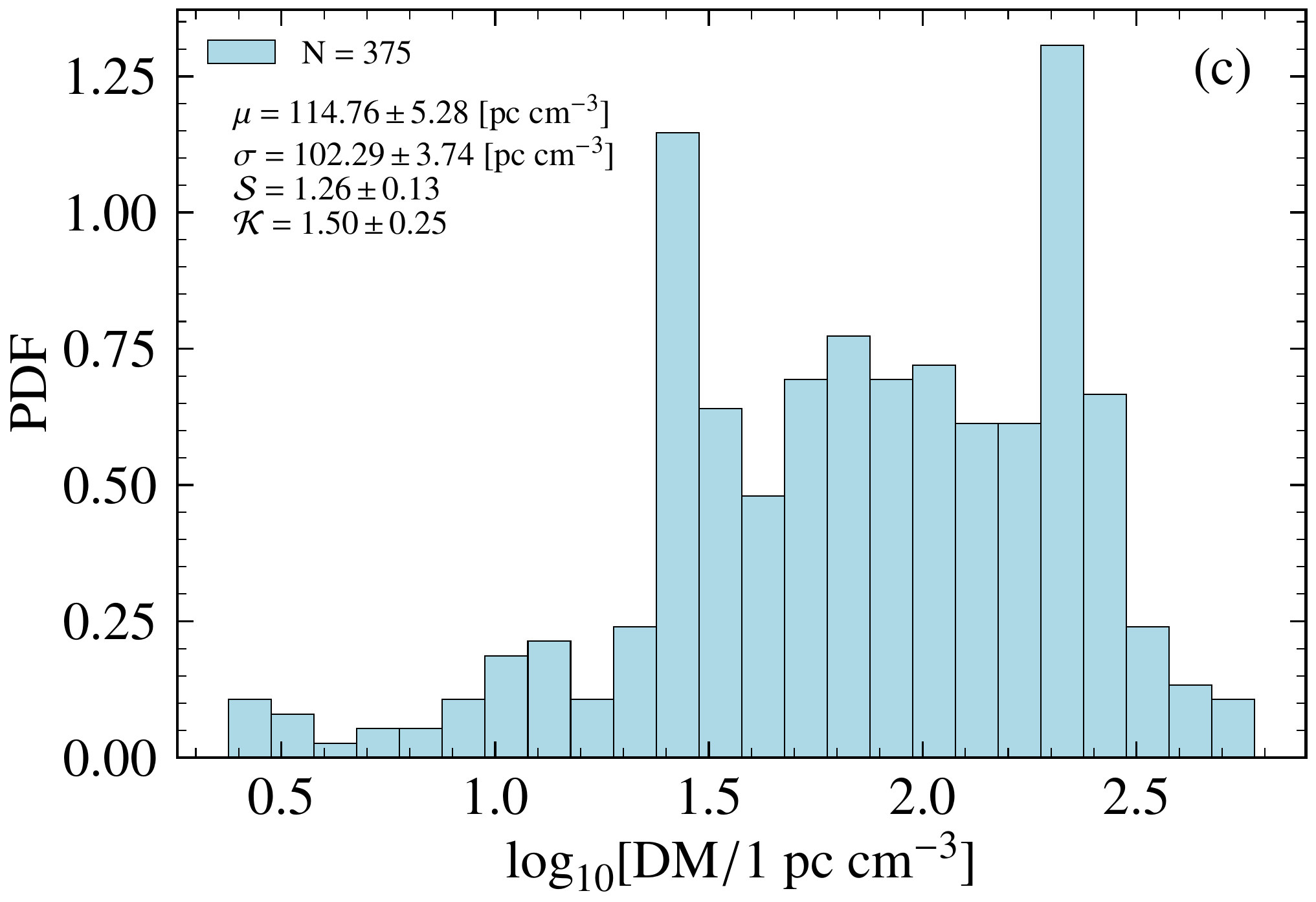}
    \end{minipage}
    \caption{Same as \Fig{data distdm} but for $\dista$. Here too, except $\RM$, both distributions show skewness ($\S$). Based on the computed kurtosis ($\kurt$), all distributions show non-normal traits (note the negative $\kurt$ for $\dista$, which is probably due to a smaller number of samples in the histogram).}
    \label{data dist a}
\end{figure*}

\Fig{data distdm}~(a) and \Fig{data dist a}~(a) present histograms for the datasets $\distdm$ and $\dista$, respectively. The mean value for $\distdm$ is measured at $(4.57 \pm 0.06)\,\kpc$, with a standard deviation of $(3.17 \pm 0.04)\,\kpc$. In contrast, $\dista$ exhibits a higher mean of $(5.12 \pm 0.17)\,\kpc$, accompanied by a larger error and a standard deviation of $(3.32 \pm 0.12)\,\kpc$. To further analyse the distributions, we calculated skewness ($\S$) and kurtosis ($\kurt$). High skewness is indicated by $|\S| > 1$, moderate skewness by $0.5 < |\S| \leq 1$, and approximate symmetry by $|\S| \leq 0.5$. A $\kurt$ near 0 suggests a Gaussian distribution, while values above 0 indicate non-Gaussian distributions with heavy tails.

For the $\distdm$ dataset, we observe moderate $\S$ and non-normal behaviour with $\kurt>0$, suggesting that the distribution deviates from normality. In contrast, the $\dista$ dataset shows an approximately symmetrical distribution ($\S < 0.5$) but also exhibits non-normal characteristics ($\kurt >0 $).

\subsubsection{$\RM$ distributions}

\Fig{data distdm}~(b) and \Fig{data dist a}~(b) show the distribution for the $\RM$s of the datasets, $\distdm$ and $\dista$, respectively. The mean value for the $\RM$ subset of $\distdm$ is measured at $(23.59 \pm 3.37)\,\rad\,\m^{-2}$, with a large standard deviation of $(124.01 \pm 2.38)\,\rad\,\m^{-2}$. In comparison, the $\dista$ dataset shows a higher mean of $(36.37 \pm 7.61)\,\rad\,\m^{-2}$ and large standard deviation of $(117.91 \pm 5.39)\,\rad\,\m^{-2}$.

Both histograms indicate minimal $\S$ (more or less symmetric distribution) and $\kurt>0$, suggesting heavy tails relative to a normal distribution. These characteristics imply that the data exhibit non-normal distribution traits. This is particularly noteworthy for $\RM$, as the observed deviations suggest that the simple random walk model \citep[see Sec. 3.2 in][]{Seta2021} may not adequately capture the complexities of the Milky Way’s magnetic field and the ISM. Instead, a more nuanced approach that considers factors such as magnetic field inhomogeneities and non-uniform electron density may be necessary for accurately describing the $\RM$ distribution. The non-normal characteristics observed likely stem from several underlying physical factors, which are discussed in \Sec{dis1}.

\subsubsection{$\DM$ distributions}
\Fig{data distdm}~(c) and \Fig{data dist a}~(c) present histograms for the $\DM$s of the datasets $\distdm$ and $\dista$, respectively. The mean value for the $\DM$ subset of $\distdm$ is measured at $(190.46 \pm 3.14)\,\pc\,\cm^{-3}$, accompanied by a very large standard deviation of $(174.05 \pm 2.22)\,\pc\,\cm^{-3}$. In contrast, the $\dista$ dataset shows a smaller mean of $(114.76 \pm 5.28)\,\pc\,\cm^{-3}$, with a correspondingly smaller standard deviation of $(102.29 \pm 3.74)\,\pc\,\cm^{-3}$.

Upon examining the distributions, we observe a significant right skewness, indicating a tendency for higher values. Moreover, the $\kurt$ values for both datasets are observed to be $> 0$, suggesting non-normal distributions. This non-normality is expected, as $\DM$ along the line of sight will be influenced by various astrophysical factors, such as variations in electron density due to different structures such as bubbles, filaments, and other features that can influence the dispersion measure, causing deviations from a normal distribution along the line of sight \citep{Ruzmaikin1988,NE2001}.

\subsection{Parameter relationships: $|\RM|$ vs $\DM$ and $\Bpa$ vs Distance}

\begin{figure*}
\includegraphics[width=\columnwidth]{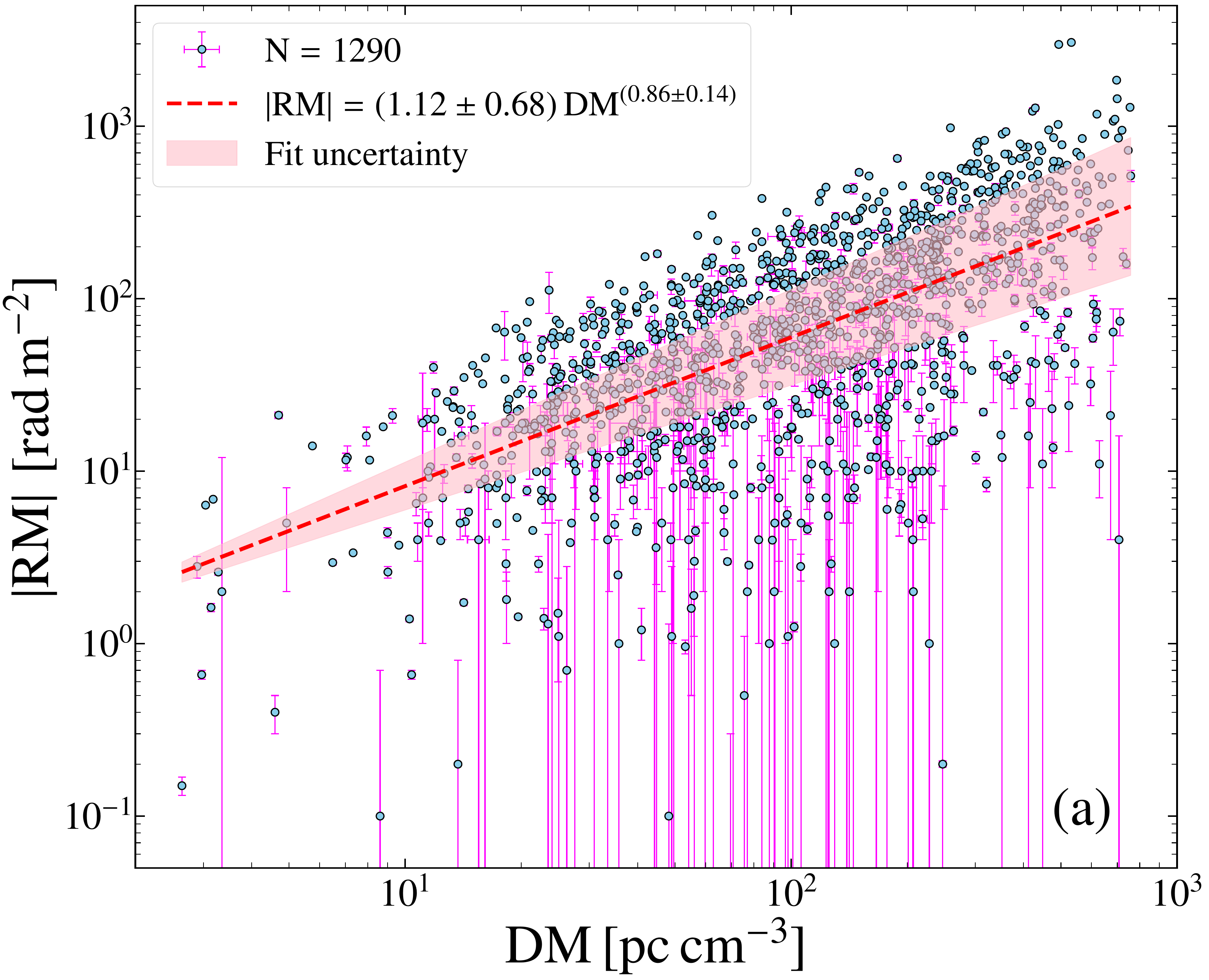} \hspace{0.05cm}
\includegraphics[width=0.96\columnwidth]{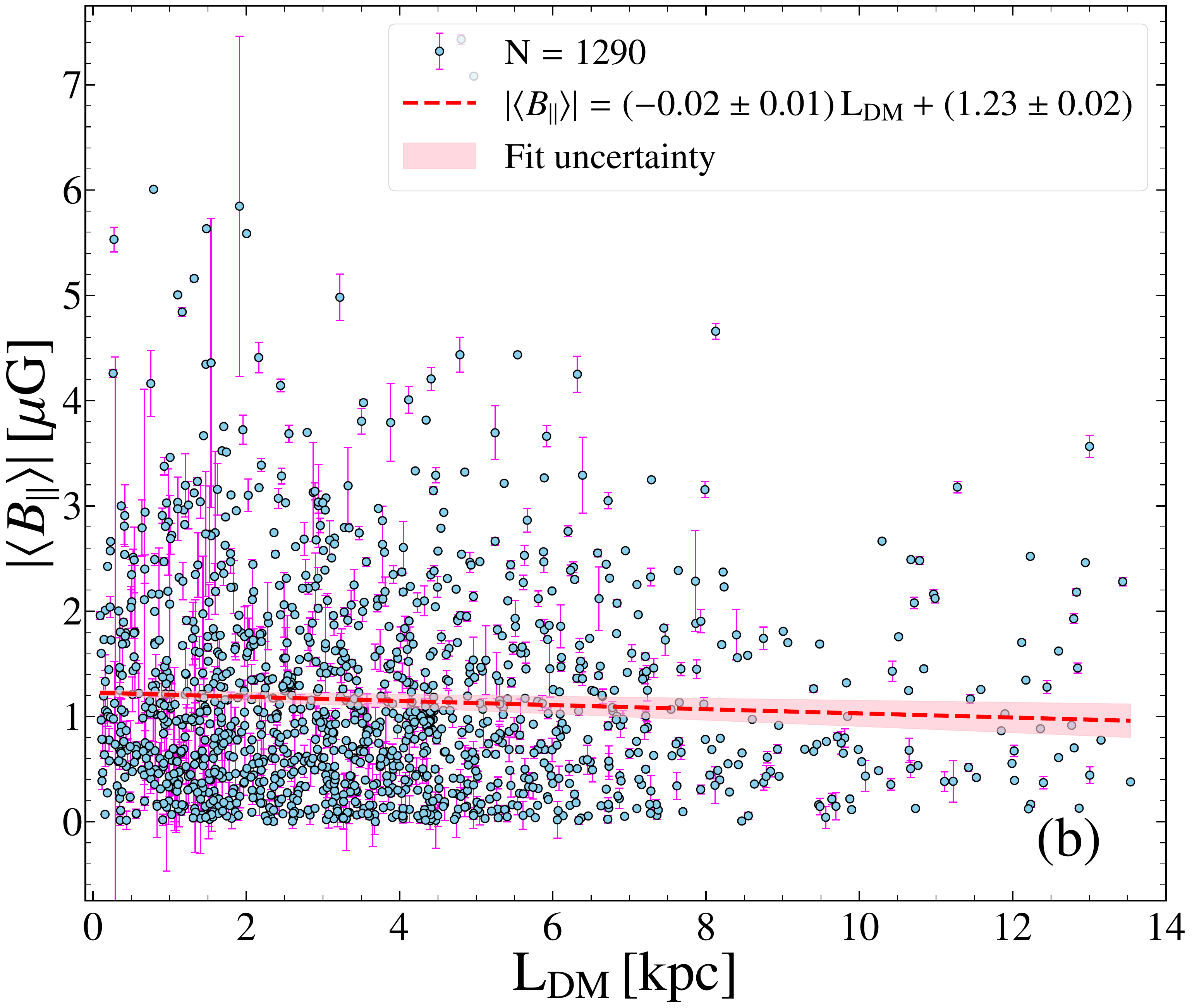}
\caption{(a): A log-log plot of $|\RM|$ versus $\DM$ for the $\distdm$ dataset, with a linear fit in the log-log space shown as the red dashed line \rev{and the one-sigma deviations are shown as a pink band around it}. The best-fit line is given by $|{\RM}| = (1.12 \pm 0.68)\,\DM^{(0.86 \pm 0.14)}$. (b) A plot of the absolute value of the average parallel magnetic field $|\langle B_{\parallel} \rangle|$ versus distance, calculated using \Eq{bp average}. The best-fit line is shown as the red dashed line \rev{with one-sigma deviations as a pink band around it}. The best-fit line is given by $|\Bpa|=(-0.02 \pm 0.01)\,\distdm + (1.23 \pm 0.02)$. From these results, we conclude that $|\RM| \propto \DM$ and $|\Bpa| \propto \distdm^0$. From (b), on an average, $|\Bpa|| \approx 1.2\,\muG$ (intercept of the fitted line).}
\label{graph4}
\end{figure*}

\begin{figure*}
\includegraphics[width=\columnwidth]{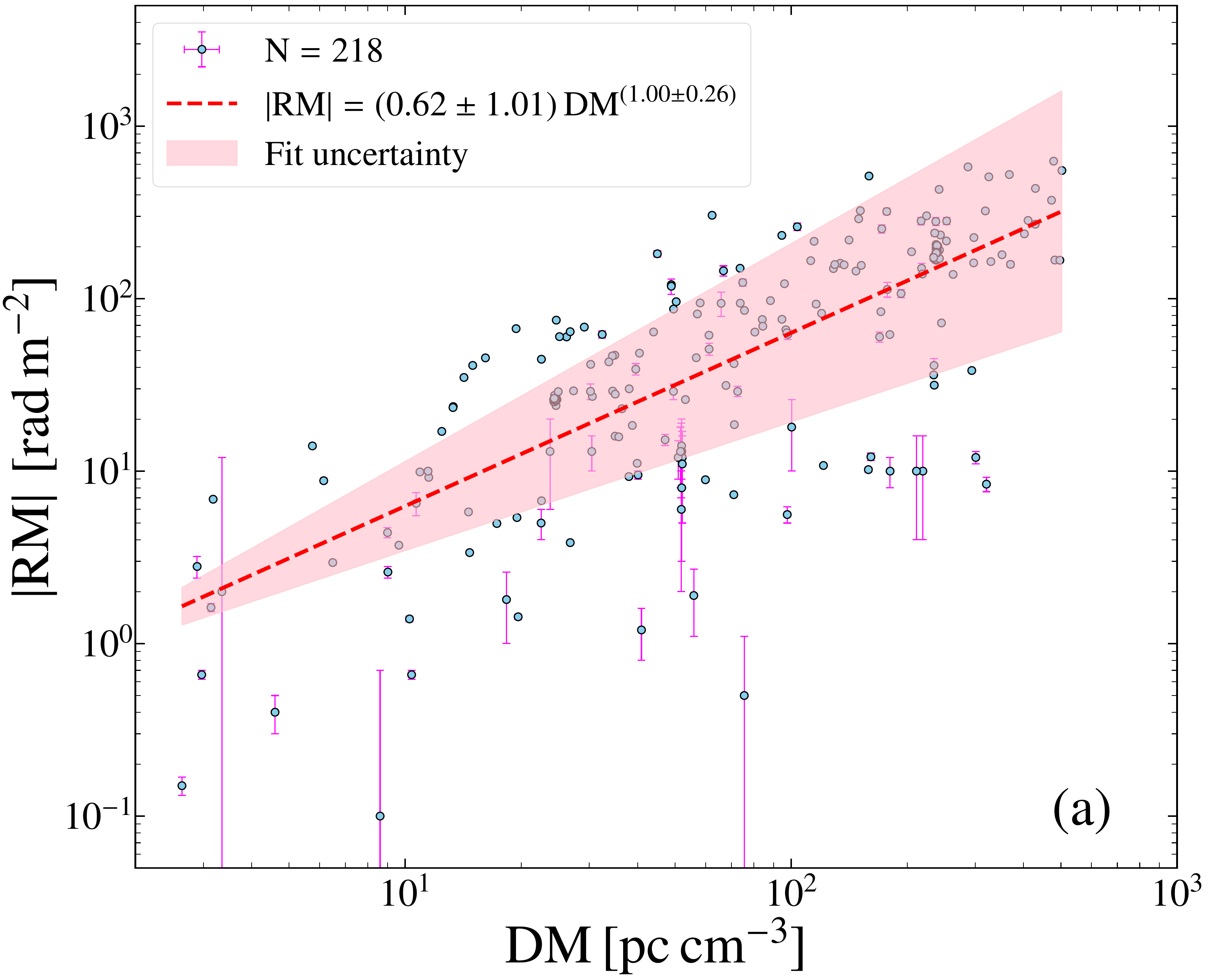} \hspace{0.05cm}
\includegraphics[width=0.96\columnwidth]{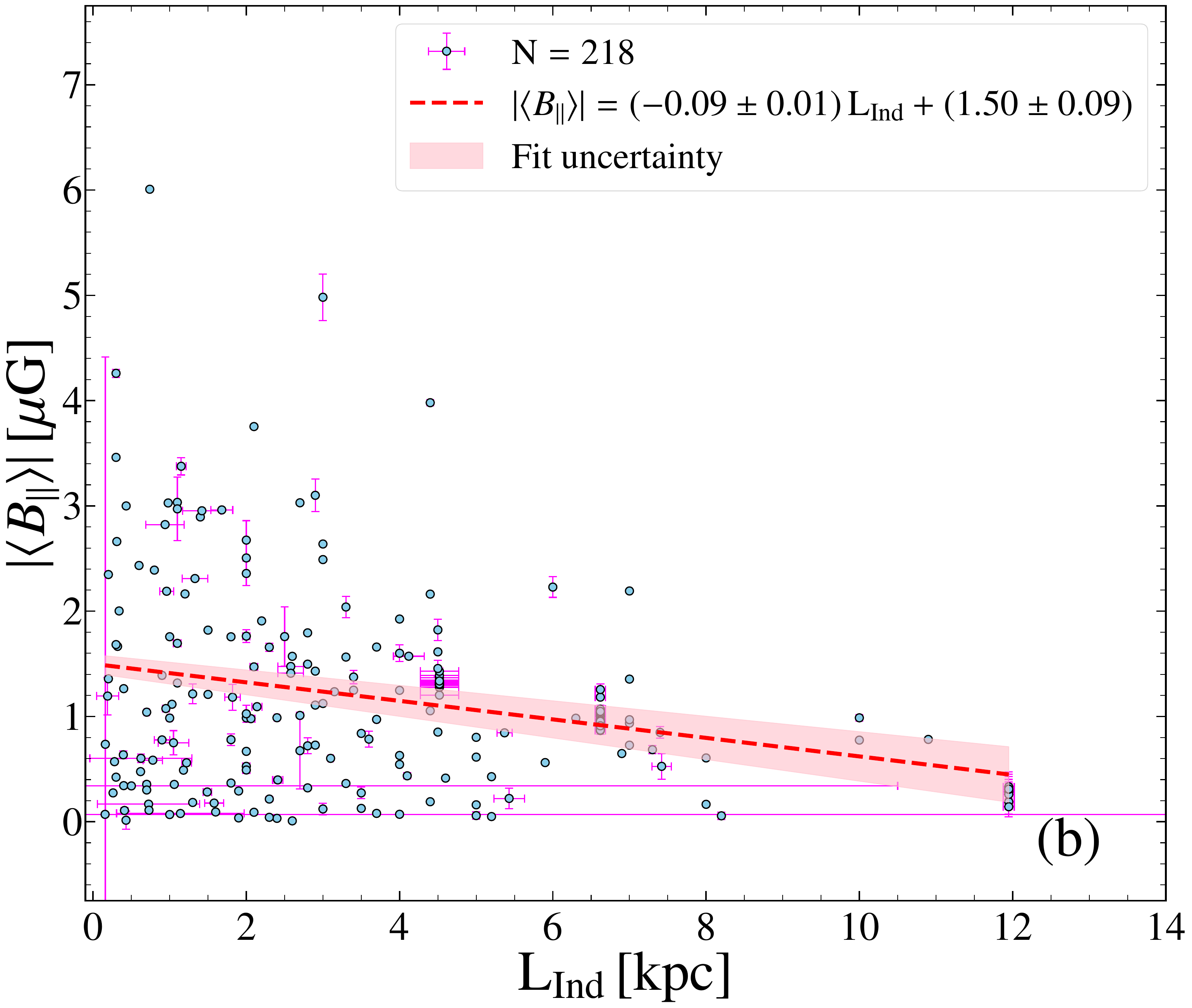}
\caption{Same as \Fig{graph4} but for the $\dista$ dataset. Here, $|{\RM}| = (0.62 \pm 1.01 )\,\DM^{(1.00 \pm 0.26)}$ and $|\Bpa| = (-0.09 \pm 0.01 )\,\dista + (1.50 \pm 0.09)$. The conclusion that $|\RM| \propto \DM$ and $|\Bpa|$ are independent of the distance to the pulsar also roughly remains the same (although the level of uncertainties is higher given the smaller size of the data). For this dataset, on average, $|\Bpa| \approx 1.5\,\muG$.}
\label{graph5}
\end{figure*}

\Fig{graph4}~(a) and \Fig{graph5}~(a) present scatter plots (blue points) showing the absolute value of $\RM$s and $\DM$s for pulsars, including their associated uncertainties. Notably, the dataset comprises $1290$ pulsars for $\distdm$ and $218$ for $\dista$.The best-fit line for the $\distdm$ dataset is given by $|{\RM}| = (1.12 \pm 0.68)\,\DM^{(0.86 \pm 0.14)}$, while for the $\dista$ dataset, it is $|{\RM}| = (0.60 \pm 0.95)\,\DM^{(1.01 \pm 0.26)}$. \rev{Both fits are consistent with a nearly linear scaling, $|{\RM}| \propto \DM$, which reflects the fact that the $\DM$–dependence of RM is primarily set by fluctuations in the thermal electron density along the line of sight. If variations in the magnetic field dominated, we would expect a weaker or more scattered correlation.}

Next, we examine the relationship between the distance to the pulsar and the estimated average magnetic field strength ($|\Bpa| = 1.232~|\RM|/\DM$) in \Fig{graph4}~(b) for $\distdm$ and \Fig{graph5}~(b) for $\dista$. The best-fit line for the $\distdm$ dataset is given by $|\Bpa|=(-0.02 \pm 0.01 )\,\dista + (1.23 \pm 0.02)$, while for the $\dista$ dataset, it is $|\Bpa|=(-0.08 \pm 0.01)\,\dista + (1.49 \pm 0.09)$. In both cases, the intercepts ($1.2\,\muG$ and $1.5\,\muG$) represent the average magnetic field strength, suggesting potentially an intrinsic baseline contribution from the large-scale Galactic field. Both slopes are close to zero, indicating no significant connection between $|\Bpa|$ and distance, suggesting that the average magnetic field strength remains consistent across the sampled pulsars, independent of the distance. However, the observed lack of significant correlation between $\Bpa$ and distance should not be taken as evidence of a uniformly strong magnetic field across all distances. Rather, this result likely stems from the averaging effect of integrating over large path lengths, which tends to smooth out smaller-scale variations in the magnetic field.

However, this somewhat `synthetic' uniformity over large $\kpc$ distances points to the importance of focusing on small-scale, localised variations within the magneto-ionic medium, where small-scale fluctuations become more prominent. These small-scale structures can often be masked by the large-scale averaging, but they play a very important role in shaping the overall behaviour of the magneto-ionic medium. This highlights the core methodology and aim of our work, to examine both the properties of the large-scale field and small-scale structures within the magneto-ionic medium. By determining the correlation lengths, $\lb$ and $\lne$, we can better understand the scale and impact of these small-scale variations, which are often masked in other analyses in the literature.

\subsection{Table of Results}

\subsection{Table of Results: $\distdm \leq 20\,\kpc$}

The $\distdm \leq 20\,\kpc$ dataset contains the largest number of pulsars and serves as the reference point for our analysis. One of the most significant findings is the difference in the correlation lengths, $\lb$ and $\lne$, across the different correlation functions. 

The magnetic field correlation length, $\lb$, ranges from $(18 \pm 8)\,\pc$ for $\Ca$ and $\Cc$ to $(22 \pm 8)\,\pc$ for $\Cb$. This indicates that small-scale magnetic fields are correlated over significantly short distances compared to the distance to the pulsar. \revst{The variation across correlation functions, though not huge, emphasises the need to choose an appropriate function to accurately capture the scale of small-scale magnetic structures.}

In contrast, the thermal electron density correlation length, $\lne$, is significantly larger. For $\Ca$, $\lne$ is $(251 \pm 12)\,\pc$, with $\Cb$ extending to $(305 \pm 12)\,\pc$, and $\Cc$ yielding $(291 \pm 12)\,\pc$. This difference, where $\lb$ ranges from $18$ to $22\,\pc$ while $\lne$ spans $251$ to $305\,\pc$, reveals that small-scale thermal electron density structures are much more extended, compared to the small-scale magnetic field structures.

The results show a fundamental difference in the scales at which magnetic fields and electron densities shape the magneto-ionic medium and the related radio observables. Magnetic fields have a more localised influence, while thermal electron density fluctuations extend across much larger scales. Recognising and distinguishing these distinct scales is crucial for accurately interpreting $\RM$ from extragalactic sources and the dynamics of the magneto-ionic medium.

Moving to other parameters, the value of $\co$ remains consistent \revst{positive} across all three correlation functions ($\Ca$, $\Cb$, and $\Cc$), averaging around $(0.05 \pm 0.02)\,\cm^{-3}\,\muG$, as shown in \Tab{results1}. \revst{This positive value suggests a dominant alignment of the magnetic field component along the line of sight.} The galactic coordinates $\gl_0$ and $\gb_0$, which serve as reference points for the large-scale magnetic field models, stabilise around $(192 \pm 9)\,\degree$ for $\gl_0$ and $(36 \pm 33)\,\degree$ for $\gb_0$, indicating that the large-scale magnetic field reference point is consistent across the correlation functions.

The average electron density, $\nea$, also remains very stable across all correlation functions for $\distdm \leq 20\,\kpc$, with values consistently around $(0.055 \pm 0.001)\,{\cm}^{-3}$, as shown in \Tab{results2}. This agrees with values in the literature \citep{GaenslerEA2008, YMW2013}.\revst{The error reflects a high degree of confidence in these measurements. This stability also likely reflects the comparatively diffuse nature of the thermal electron density in the ISM.} \rev{The small error bars represent formal statistical uncertainties and do not include systematic effects from model assumptions, though the similar estimate with different distances may still imply a converged solution for the magnitude of the large-scale thermal electron density.}

The calculated large-scale magnetic field strength, $|B|$, for $\distdm \leq 20\,\kpc$ remains consistent across all correlation functions, with a value of approximately $1.2 \pm 0.5\,\muG$ for $\Ca$, $\Cb$, and $\Cc$. \revst{The positive sign of $B$ indicates that the magnetic field is primarily aligned along the line of sight.} It should be noted that $B$ has large error bars, which is largely due to uncertainties in the parameter $\co$, which stems from the relatively high uncertainty in $\RM$ observations compared to $\DM$.

\begin{table*}
\caption{Summary of results for various datasets categorised by the distance condition, including $\distdm \leq 1\,\kpc$, $\distdm \leq 2\,\kpc$, $\distdm \leq 5\,\kpc$, $\distdm \leq 20\,\kpc$, and $\dista$. Each distance condition is analysed using three correlation functions, $\Ca$, $\Cb$, and $\Cc$. The table presents the number of data points ($\ndata$) and the derived parameters, including $\co$, $\gl_0$, $\gb_0$, and $\lb$, along with their associated uncertainties. The parameters $\co$, $\gl_0$, $\gb_0$, and $\lb$ exhibit slight variations with distance, particularly a decrease in $\co$ with increasing distance (up to 5 $\kpc$) and an increase in $\lb$ over the same range is observed. The results for the $\dista$ dataset are significantly different from others (except for the $\distdm \leq 1\,\kpc$ case, where they are more similar). Additionally, within uncertainities, $\Ca$, $\Cb$, and $\Cc$ give similar $\lb$ values across distances.}
\centering
{
\begin{tabular}{lcccccc}
\hline 
Data Set & $\ndata$ & {$C(s)$} & {$\co[\cm^{-3}\,\muG]$} & {$\gl_0[\degree]$} & {$\gb_0[\degree]$} & {$\lb[\pc]$} \\ \hline
$\distdm \leq 1\,\kpc$ & 165 & $\Ca$ & 0.06 $\pm$ 0.03 & 124 $\pm$ 29 & 51 $\pm$ 19 & 11 $\pm$ 5\\
$\distdm \leq 1\,\kpc$ & 165 & $\Cb$ & 0.06 $\pm$ 0.03 & 124 $\pm$ 29 & 50 $\pm$ 19 & \rev{14} $\pm$ 5 \\
$\distdm \leq 1\,\kpc$ & 165 & $\Cc$ & 0.06 $\pm$ 0.03 & 125 $\pm$ 31 & 54 $\pm$ 19 & 12 $\pm$ 5 \\
$\distdm \leq 2\,\kpc$ & 397 & $\Ca$ & 0.04 $\pm$ 0.01 & 119 $\pm$ 27 & 65 $\pm$ 12 & 35 $\pm$ 14 \\
$\distdm \leq 2\,\kpc$ & 397 & $\Cb$ & 0.04 $\pm$ 0.01 & 119 $\pm$ 26 & 65 $\pm$ 12 & \rev{44} $\pm$ 14 \\
$\distdm \leq 2\,\kpc$ & 397 & $\Cc$ & 0.04 $\pm$ 0.01 & 111 $\pm$ 28 & 66 $\pm$ 13 & 40 $\pm$ 14 \\
$\distdm \leq 5\,\kpc$ & 936 & $\Ca$ & 0.05 $\pm$ 0.01 & 122 $\pm$ 10 & 66 $\pm$ 10 & 47 $\pm$ 17 \\
$\distdm \leq 5\,\kpc$ & 936 & $\Cb$ & 0.05 $\pm$ 0.01 & 123 $\pm$ 10 & 65 $\pm$ 10 & \rev{57} $\pm$ 17 \\
$\distdm \leq 5\,\kpc$ & 936 & $\Cc$ & 0.04 $\pm$ 0.01 & 125 $\pm$ 11 & 65 $\pm$ 11 & 51 $\pm$ 17 \\
$\distdm \leq 20\,\kpc$ & 1290 & $\Ca$ & 0.05 $\pm$ 0.02 & 192 $\pm$ 9 & 36 $\pm$ 33 & 18 $\pm$ 8 \\
$\distdm \leq 20\,\kpc$ & 1290 & $\Cb$ & 0.05 $\pm$ 0.02 & 192 $\pm$ 9 & 36 $\pm$ 33 & \rev{22} $\pm$ 8 \\
$\distdm \leq 20\,\kpc$ & 1290 & $\Cc$ & 0.05 $\pm$ 0.02 & 192 $\pm$ 9 & 36 $\pm$ 34 & 18 $\pm$ 8 \\
$\dista$ & 218 & $\Ca$ & 0.07 $\pm$ 0.02 & 133 $\pm$ 9 & 18 $\pm$ 30 & 4 $\pm$ 1 \\
$\dista$ & 218 & $\Cb$ & 0.07 $\pm$ 0.02 & 134 $\pm$ 9 & 18 $\pm$ 31 & \rev{5} $\pm$ 1 \\
$\dista$ & 218 & $\Cc$ & 0.07 $\pm$ 0.02 & 134 $\pm$ 9 & 18 $\pm$ 30 & 4 $\pm$ 1 \\
\hline
\end{tabular}
}
\label{results1}
\end{table*}

\begin{table*}
\caption{The results and errors for $\nea$ and $\lne$, categorised by the distance conditions: $\distdm \leq 1\,\kpc$, $\distdm \leq 2\,\kpc$, $\distdm \leq 5\,\kpc$, $\distdm \leq 20\,\kpc$, and $\dista$. Similar to \Tab{results1}, the parameter $\nea$ remains consistent across correlation functions, while $\lne$ shows greater variation. Notably, $\nea$ increases with distance up to 5 $\kpc$. As in \Tab{results1}, the results for the $\dista$ dataset are significantly different  except for the $\distdm \leq 1\,\kpc$ case.}
\centering
\begin{tabular}{lcccccc}
\hline
Data Set & $\ndata$ &{$C(s)$} & $\nea[\cm^{-3}]$ & $\lne[\pc]$ \\ \hline
$\distdm  \leq 1\,\kpc$ & 268 & $\Ca$ & 0.106 $\pm$ 0.004 & 31 $\pm$ 10 \\
$\distdm  \leq 1\,\kpc$ & 268 & $\Cb$ & 0.106 $\pm$ 0.004 & \rev{37} $\pm$ 10 \\
$\distdm  \leq 1\,\kpc$ & 268 & $\Cc$ & 0.104 $\pm$ 0.004 & 34 $\pm$ 10 \\
$\distdm  \leq 2\,\kpc$ & 693 & $\Ca$ & 0.069 $\pm$ 0.002 & 89 $\pm$ 10 \\
$\distdm  \leq 2\,\kpc$ & 693 & $\Cb$ & 0.069 $\pm$ 0.002 & \rev{114} $\pm$ 10 \\
$\distdm  \leq 2\,\kpc$ & 693 & $\Cc$ & 0.063 $\pm$ 0.002 & 115 $ \pm$ 10 \\
$\distdm  \leq 5\,\kpc$ & 1898 & $\Ca$ & \rev{0.060} $\pm$ 0.001 & \rev{178} $\pm$ \rev{4} \\
$\distdm  \leq 5\,\kpc$ & 1898 & $\Cb$ & \rev{0.060} $\pm$ 0.001 & \rev{250} $\pm$ \rev{4} \\
$\distdm  \leq 5\,\kpc$ & 1898 & $\Cc$ & 0.057 $\pm$ 0.001 & \rev{252} $\pm$ \rev{4} \\
$\distdm  \leq 20\,\kpc$ & 3077 & $\Ca$ & 0.055 $\pm$ 0.001 & \rev{251} $ \pm$ 12 \\
$\distdm  \leq 20\,\kpc$ & 3077 & $\Cb$ & \rev{0.054} $\pm$ 0.001 & \rev{305} $ \pm$ 12 \\
$\distdm  \leq 20\,\kpc$ & 3077 & $\Cc$ & \rev{0.053} $\pm$ 0.001 & \rev{291} $ \pm$ 12 \\
$\dista$ & 375 & $\Ca$ & 0.033 $\pm$ 0.002 & \rev{35} $\pm$ 15 \\
$\dista$ & 375 & $\Cb$ & 0.033 $\pm$ 0.002 & \rev{60} $\pm$ 15 \\
$\dista$ & 375 & $\Cc$ & 0.033 $\pm$ 0.002 & \rev{39} $\pm$ 15 \\ \hline
\end{tabular}
\label{results2}
\end{table*}

\begin{table*} 
\caption{Large-scale magnetic field strength, $|B|$, along with their determined uncertainties for distance conditions: $\distdm \leq 1\,\kpc$, $\distdm \leq 2\,\kpc$, $\distdm \leq 5\,\kpc$, $\distdm \leq 20\,\kpc$, and $\dista$, and are further divided into three correlation functions, $\Ca$, $\Cb$, and $\Cc$. The form of the correlation function does not play a major role in the derived large-scale magnetic field strengths. Notably, $|B|$ tends to increase slightly with distance, and the values for the $\dista$ dataset are higher than those for $\distdm$. This discrepancy suggests a potential bias in the sample with independently determined pulsar distances.}
\centering
\begin{tabular}{lcc}
\hline
Data Set &{$C(s)$} & $|B| [\muG]$ \\ \hline
$\distdm \leq 1\,\kpc$ & $C_1$ & 0.7 $\pm$ 0.3 \\
$\distdm \leq 1\,\kpc$ & $C_2$ & 0.7 $\pm$ 0.3 \\
$\distdm \leq 1\,\kpc$ & $C_3$ & 0.7 $\pm$ 0.3 \\
$\distdm \leq 2\,\kpc$ & $C_1$ & 0.8 $\pm$ 0.3 \\
$\distdm \leq 2\,\kpc$ & $C_2$ & 0.8 $\pm$ 0.3 \\
$\distdm \leq 2\,\kpc$ & $C_3$ & 0.8 $\pm$ 0.3 \\
$\distdm \leq 5\,\kpc$ & $C_1$ & 0.9 $\pm$ 0.3 \\
$\distdm \leq 5\,\kpc$ & $C_2$ & 0.9 $\pm$ 0.3 \\
$\distdm \leq 5\,\kpc$ & $C_3$ & 0.9 $\pm$ 0.3 \\
$\distdm \leq 20\,\kpc$ & $C_1$ & 1.2 $\pm$ 0.5 \\
$\distdm \leq 20\,\kpc$ & $C_2$ & 1.2 $\pm$ 0.5 \\
$\distdm \leq 20\,\kpc$ & $C_3$ & 1.2 $\pm$ 0.5 \\
$\dista$ & $C_1$ & 2.7 $\pm$ 0.6 \\
$\dista$ & $C_2$ & 2.5 $\pm$ 0.6 \\
$\dista$ & $C_3$ & 2.6 $\pm$ 0.6 \\ \hline
\end{tabular}
\label{results3}
\end{table*}
The main results obtained from our study are given in \Tab{results1}, \Tab{results2}, and \Tab{results3}. Here, we discuss the interesting trends and important points from those results.

\subsection{Accounting for non-Gaussianity in $\RM$ and $\DM$ distributions} \label{sec:nonGaussian}

\rev{
\Fig{data distdm} and \Fig{data dist a} show that both the $\RM$ and $\DM$ distributions are non-Gaussian, with kurtosis ($\kurt$) significantly greater than zero. To assess the sensitivity of our results to this non-Gaussianity, we draw $\epsilon_{\RM}$ in \Eq{rm4} and $\eta_{\DM}$ in \Eq{dm4} from random, non-Gaussian distributions (also, always making sure that $\DM > 0\,\pc\,\cm^{-3}$). These non-Gaussian distributions are generated using the Fleishman transformation \citep{Fleishman1978}, which allows us to set the skewness and kurtosis while preserving the empirical mean and standard deviation of the data. For both $\RM$ and $\DM$, we set the skewness to zero, consistent with the near-symmetric observed distributions, and tested our method across a range of kurtosis values ($\kappa = 0, 0.1, 1, 2, 5$). For each $\kappa$, we repeated the full parameter estimation procedure outlined in Sections~\Sec{mainmethod1} -- \Sec{mainmethod4}.

We find that the fitted parameters, in particular $\co$, $\lb$, and $\lne$, remain consistent with those obtained under the Gaussian assumption, within their respective uncertainties. This demonstrates that our results are robust to deviations from Gaussianity in the $\RM$ and $\DM$ distributions. The results for the non-Gaussian distributions are presented and further discussed in \App{app3}.
}

\subsection{Impact of distance on $\co$, $\gl_0$, $\gb_0$, and $\lb$}\label{3.5}

Originally, we wanted to study how different parts of the Milky Way influence the properties of the magnetic field and thermal electron density, especially $\lb$ and $\lne$. Focusing our attention mainly on the Galactic disk, where most of the gas and star-forming activity is expected, we tried to isolate these effects by limiting the dataset only to $\gb$ values between $-5^\circ$ and $+5^\circ$. However, with a limited number of pulsars \rev{($\ndata = 120$)} in this range, we were unable to obtain statistically robust fits.

\rev{We also attempted to separate the pulsar sample into longitude-based subsamples: one roughly aligned with the local spiral arms ($85^\circ \leq \gl \leq 95^\circ$ and $265^\circ \leq \gl \leq 275^\circ$), which should predominantly probe the local large-scale magnetic field and one in the perpendicular direction ($355^\circ \leq \gl \leq 5^\circ$ and $175^\circ \leq \gl \leq 185^\circ$), which should be more sensitive to the random component of the magnetic field. Unfortunately, the available number of pulsars in these restricted regions ($\ndata = 49$ and $\ndata = 115$, respectively) was too small to obtain statistically converged results.}

\rev{We therefore focused in this paper on the broader distance-limited samples (1, 2, 5, and 20 $\kpc$), which provide statistically stable and converged estimates of the magnetic field and electron density parameters. Nonetheless, the subsample analysis highlights the limitations of current datasets in probing localised magnetic field structures. Future surveys, particularly with SKA \citep{KeaneEA2015}, are expected to provide much larger and more uniformly distributed pulsar samples, at which point such region-specific analyses will become feasible and yield more localised constraints on the Galactic magnetic field.}

The results in \Tab{results1} reveal an interesting trend in the magnetic field parameters as a function of the cutoff distance to the pulsars. The correlation length $\lb$ shows variation with distance, suggesting the influence of different ISM regions on the correlation length. For $\distdm \leq 2\,\kpc$, $\lb$ values increase, reaching $(35 \pm 14)\,\pc$ for $\Ca$, $(44\pm 14)\,\pc$ for $\Cb$, and $(40 \pm 14)\,\pc$ for $\Cc$. As the distance extends to $\distdm \leq 5\,\kpc$, $\lb$ values continue to increase, with $\lb$ reaching $(47 \pm 16)\,\pc$ for $\Ca$, $(57 \pm 16)\,\pc$ for $\Cb$, and $(51 \pm 16)\,\pc$ for $\Cc$. These larger $\lb$ values may indicate regions where the magnetic field is more coherent and less random, suggesting that the ISM’s large-scale structure becomes more prominent as the sight line extends through different ISM environments, including more quiescent regions or spiral arms. A caveat could also be that with a decreasing number of pulsars in the dataset with smaller distances ($\ndata$), the separation between the small- and large-scale might not be as good, potentially contaminating the estimated $\lb$. \revb{Furthermore, the significant differences in some parameters for $\distdm \leq 5\,\kpc$ and $\distdm \leq 20\,\kpc$ may be due to contributions from the known large-scale magnetic field reversal, which affects the $\RM$s of distant pulsars. In addition, some of these pulsars probe distant regions within the same spiral arm, where the line-of-sight magnetic field could be opposite to that in the nearby part of the arm.}

\revst{
At shorter distances ($\distdm \leq 1\,\kpc$), the parameter $\co$ is consistently, around $(0.06 \pm 0.03)\,\cm^{-3}\,\muG$ across all three correlation functions, indicating a possible reversal or variation in the magnetic field alignment due to localised structures. As the distance increases to $\distdm \leq 2\,\kpc$, $\co$ remains negative but with smaller magnitudes, around $(0.04 \pm 0.01)\,\cm^{-3}\,\muG$ for $\Ca$,$\Cb$, and $\Cc$, suggesting a somewhat of a gradual decrease in the magnetic field. By $\distdm \leq 5\,\kpc$, $\co$ increases ever so slighty at $(0.05 \pm 0.01)\,\cm^{-3}\,\muG$ for $\Ca$ and $\Cb$ and $(0.04 \pm 0.02)\,\cm^{-3}\,\muG$ for $\Cc$. However, the large error bars, particularly at shorter distances, highlight the significant uncertainties associated with $\co$. These uncertainties likely arise from the complex and multiscale nature of the Galactic magnetic fields, making it harder to pin down the sign of $\co$, especially at closer ranges with smaller size of datasets.}

\rev{
We note that although some inferred values of $\gb_0$ in Table 1 appear large, their uncertainties are substantial, particularly for larger distance cuts (e.g., $\distdm \leq 20,\kpc$). Within these uncertainties, the results remain consistent with $\gb_0$ close to $0^{^{\circ}}$, in agreement with the established view that the Galactic magnetic field lies largely parallel to the plane \citep{Han2017}. At smaller distances, the scatter in $\gb_0$ reflects the reduced number of pulsars and limited line-of-sight coverage, and we therefore interpret the apparent variations as sampling effects rather than evidence of a significant vertical field.

The inferred $\gl_0$ values also differ somewhat from earlier pulsar- and extragalactic source-based studies. These shifts primarily occur because our analysis simultaneously models both large- and small-scale components using a larger pulsar dataset, which can alter the best-fit orientation. Nonetheless, our results are broadly consistent within uncertainties, complementing earlier estimates and highlighting the influence of different datasets and modelling approaches.
}

\subsection{Impact of distance on $\nea$, $\lne$, and $|B|$}
The correlation length $\lne$ shows significant variability with distance \revst{, particularly when using the $\Cb$ correlation function, which is slightly more sensitive to small changes in electron density scales}. At shorter distances ($\distdm\leq 1\,\kpc$), $\lne$ values range from $31$ to $37\,\pc$, indicating the influence of smaller-scale structures. However, as the distance extends to $\distdm \leq 5\,\kpc$, $\lne$ increases dramatically to over $252\,\pc$, suggesting that the magneto-ionic structures become more correlated over larger spatial scales at greater distances. \revst{The sensitivity of $\Cb$ highlights the need to choose the appropriate correlation function based on the scale of analysis for accurate insights into electron density variations. This is further explored in ....}

In contrast, the average electron density $\nea$, remains relatively stable across different distances \revst{, with small error bars indicating reliable estimates}. At shorter distances ($\distdm \leq 1\,\kpc$), $\nea$ is around $0.106 \pm 0.004\,{\cm}^{-3}$, gradually decreasing to $0.060 \pm 0.001\,{\cm}^{-3}$ by $\distdm \leq 5\,\kpc$. This steady decline reflects the more diffuse nature of the magneto-ionic medium at greater distances.

Magnetic field strength, $|B|$, derived from $\co$ and $\nea$, remains consistent across the sampled distances, averaging around $1.2 \pm 0.5\,\muG$ up to $\distdm \leq 20\,\kpc$. \revst{At closer distances ($\distdm \leq 1\,\kpc$), $B$ shows slight negative values, potentially indicating reversals in the large-scale magnetic field. As distance increases, $B$ shifts slightly positively, suggesting that the large-scale magnetic field aligns more with the line of sight. However, given the moderate error bars, these variations are not statistically robust.} \rev{Although, for all cases, the uncertainties are significant ($\approx 30$ -- $40\%$) but this remains true across all the sampled distances for the $\distdm$ dataset. The value of $\approx 1\,\muG$ is broadly consistent with that inferred from polarised synchrotron emission observations \citep{Beck2016}.}

\subsection{Dataset variations: $\distdm$ vs $\dista$}
The $\dista$ dataset, which utilises independently obtained pulsar distances (distance estimated without taking into account the pulsar $\DM$s), shows significant differences compared to the $\distdm$ dataset. In particular, the magnetic correlation length, $\lb$, is significantly smaller ($4 \,\pc$ -- $5\,\pc$), indicating that really local fluctuations dominate this dataset, possibly due to selection biases such as `clustered' pulsars in globular clusters (see bunches at the same distances in \Fig{graph5}~(b)).

Both $\nea$ and $\lne$ are lower in the $\dista$ dataset compared to $\distdm$. $\nea$ hovers around $(0.033 \pm 0.002)\,{\cm}^{-3}$, while $\lne$ ranges from $35\,\pc$ to $60\,\pc$, reflecting the more localised structures sampled in $\dista$. These lower values suggest that this dataset captures different aspects of the magneto-ionic medium, likely more influenced by the local ISM environments.

The $|B|$ values in the $\dista$ dataset are also higher, ranging from $(2.5 \pm 0.6)\,\muG$ to $(2.7 \pm 0.6)\,\muG$. These larger values may result from the smaller $\lb$ and $\lne$ in these environments, which lead to stronger magnetic field estimates when averaging over shorter distances. However, it’s important to note that the error bars are larger in the $\dista$ dataset due to its significantly smaller sample size. This further underscores the influence of environmental factors and potential selection biases, which are further explored in \Sec{pulsfactors}.

\revst{By contrast, the $\distdm$ dataset relies on the thermal electron density model, which estimates pulsar distances indirectly via the observed $\DM$. While this method is practical due to the larger sample size, it introduces potential biases due to its dependence on electron density models (which are derived from $\DM$ values). The large inherent differences between the two datasets add complexity to interpreting the differences in the obtained results.}

\section{Discussion} \label{sec:dis}

\subsection{Difference between $\lb$ and $\lne$}\label{disdiv}

The stark contrast between $\lb$ and $\lne$ highlights that different physical processes within the Milky Way’s magneto-ionic medium control their properties. While $\RM$ traces the product of thermal electron density and line-of-sight magnetic field strength, $\DM$ reflects the total electron column density \citep[e.g.~see Section 2 of][]{Hutschenreuter2023}. Our results show that $\lb$ is much smaller (around $20$ -- $30\,\pc$) compared to $\lne$, which spans $250$ -- $300\,\pc$. This indicates that magnetic field fluctuations occur on much shorter spatial scales, whereas electron density variations persist over much larger scales. This also suggests that magnetic fields and electron densities \rev{probably} respond to different dominant physical mechanisms.

Regions of high turbulence, such as those affected by star formation, supernovae, or spiral arm shocks, experience intense dynamical processes. These events disrupt magnetic field lines, leading to tangling and compression that reduce magnetic coherence, reflected in smaller $\lb$ values \citep{Ferriere2020, Ricarte, SetaEA2020, SetaF2021dyn}. \rev{In contrast, beyond turbulence, small-scale electron density structures are likely shaped by one or more large-scale HII regions \citep[sizes in the range $10$ -- $200\,\pc$, see][]{Azimlu}, which ionise the surrounding medium, probably explaining the larger $\lne$.}

\rev{Small-scale magnetic fields in the ISM are expected to be strongly influenced by turbulence and dynamo action \citep{Rincon2019}. Numerical simulations at modest values of the magnetic Reynolds number ($\mathrm{Rm}$ of the order of $10^{3}$) produce magnetic correlation scales that are a fraction ($1/2$ -- $1/3$) of the turbulence driving scale \citep[e.g.][]{SchekochihinEA2004, HaugenEA2004, SetaEA2020, SetaF2021dyn}. Given the realistic ISM conditions, however, $\mathrm{Rm}$ is expected to be much larger  \citep[$\approx 10^{18}$, see Table 1 in][]{BrandenburgS2005}. The final state of the dynamo-generated magnetic fields at such high $\mathrm{Rm}$s remains unsettled but analytical models suggest that its properties approach those at the critical value of $\mathrm{Rm}$ for the dynamo action \citep[$\approx 10^{2}$, see Table III in][]{SetaEA2020} due to strong non-linearity \citep{subramanian1999, subramanian2003}. Assuming this is the case and the turbulence is driven by supernova explosions at $\ell_{\rm driv} \approx 100\,\pc$, the estimated values of $\lb$ (20 -- 30 $\pc$) are consistent with the turbulent dynamo action at smaller scales. We also note that some dynamo theories predict a final magnetic correlation scale as $\approx \ell_{\rm driv} \, \mathrm{Rm}^{-1/2}$ \citep[see][]{Dittrikh1988} and this would yield an unrealistically small magnetic field correlation scale.}

This distinction between $\lb$ and $\lne$ is crucial for interpreting $\RM$ data, particularly from extragalactic sources where $\DM$ is often unavailable (except for FRBs). While extragalactic $\RM$s provide broad sky coverage, they lack the $\ne$ information that pulsar $\DM$s offer, complicating the separation of magnetic and electron density contributions \citep{Oppermann2012, Hutschenreuter2020, Hutschenreuter2022, VanEck2023}. Future FRB datasets, which directly provide $\DM$s, will help bridge this gap and improve the joint use of $\RM$ and $\DM$ to probe both the Galactic and extragalactic magneto-ionic media \citep{Lorimer2007, RaviEA2016}.

\subsection{Non-Gaussian $\RM$ distribution}\label{dis1}

The non-normal distribution of $\RM$, shown in \Fig{data distdm}~(b) and \Fig{data dist a}~(b), likely arises from multiple factors:
 \begin{enumerate}
     \item Magnetic field structure: The Milky Way’s magnetic field includes both large-scale, regular fields, such as those in spiral arms \citep{BeckEA2019, MaEA2020}, and small-scale random fields. \rev{A simple combination of a large-scale field with Gaussian random fluctuations would result in a shifted Gaussian distribution, i.e., a Gaussian with a non-zero mean. But this combination at different levels for different regions within the Milky Way, especially with arm and interarm contrast \citep{Shukurov1998}, might introduce non-Gaussianity.}
     \item Non-uniform electron density: The ionised gas in the Milky Way varies due to localised ionisation sources and the multiphase ISM \citep{Ha2023}. Variations from HII regions, supernova remnants, and enhanced star formation can add complexity to the $\RM$ distribution.
     \rev{
     \item Magnetic field - thermal electron density correlation: Throughout our method, we assume that the magnetic field and thermal electron density are uncorrelated. However, on much smaller sub-kpc scales, such a correlation could probably exist due to shock compression. This makes the $\RM$ distribution significantly non-Gaussian \citep[this is observed in $\RM$ distributions derived from turbulent magnetohydrodynamic simulations, e.g.~see Fig.~3(a) in][]{Seta2021}
     }
     \item ISM inhomogeneities: The inhomogeneous ISM, with structures like filaments, bubbles, and voids, further complicates the $\RM$ distribution \citep{Martizzi2015}.
     \rev{
     \item Sample inhomogeneities: The pulsar sample is far from uniform (see \Fig{spirals}) and thus such a non-uniform sampling might introduce significant deviations from Gaussianity.
     }
 \end{enumerate}
Finding the exact reason for the observed non-Gaussianity in $\RM$ distributions requires further work and a combination of observational data, numerical simulations, and analytical models \citep[especially with regards to \rev{other relevant observations, in particular HI and H$\alpha$ observations}, see][]{BoulangerEA2018}.

\subsection{Considering pulsars as probes of the Galactic magnetic fields}\label{pulsfactors}

Pulsars provide a valuable tool for probing the Milky Way’s magnetic field, though their use comes with certain limitations due to environmental factors, line-of-sight averaging, uneven spatial distribution, and uncertainties in distance estimates. Often located in turbulent environments like supernova remnants (SNRs) and star-forming regions, pulsars are associated with high-density ionised gas and amplified local magnetic fields, which can skew $\RM$ and $\DM$ measurements, making them less representative of the broader magneto-ionic medium \citep{Ferriere2001, HaverkornEA2008, GaenslerEA2008}.

Additionally, pulsars are clustered around the Galactic Center, spiral arms, and globular clusters, where star formation is active and magnetic fields are stronger, resulting in a geographic bias toward higher $\RM$ and $\DM$ values \citep{ManchesterEA2005}. This uneven distribution may lead to overestimations in parameters like $\nea$ and $|B|$, while quieter, less active regions are undersampled, complicating generalisations about the Galactic magnetic field.

Additionally, scattering effects in dense ISM regions broaden pulsar signals, distorting $\RM$ and $\DM$ measurements \citep[this distortion can also be used to study the intervening ISM properties, see][]{SobeyEA2021, OswaldEA2021} and masking small-scale magnetic variations through line-of-sight averaging. Together, these effects smooth out local fluctuations, presenting a more uniform but potentially misleading picture of the magnetic field.

Pulsar distances are often estimated using electron density models \citep{NE2001, YMW2013}, which can have considerable uncertainties, especially in complex regions \citep[e.g.~see][]{KoljonenEA2024}. The independent distance in $\dista$ dataset does address this issue partially, but the current small data size, combined with probable biases, makes it difficult to fully utilise their potential and also compare the results with the $\distdm$ dataset. 

\subsection{\rev{Assumptions and related caveats}} \label{4.4}

\rev{
Throughout our analysis, driven by the analytical nature of the work and the limitations of the available data, we have made some simplifying assumptions. Here, we discuss the possible implications of the major assumptions in the study and the related caveats.

Our large-scale magnetic field model, \Eq{RM_l}, assumes that the field is unidirectional and does not capture the complexities of the Galactic magnetic fields, especially the details of the spiral structure \revb{and the presence of magnetic field reversals}. Several models in the literature include details of the large-scale magnetic field structure \citep{SunEA2008, VanEckEA2011, Jansson2012, shukurov2019}. Unlike our simplistic model, such models have a large number of empirical or physical parameters for which the current data are severely limited (\Fig{spirals}) and these models also do not agree with each other \cite[e.g.~see Fig.~8 in][]{MaEA2020}. Thus, some of the $\RM$ fluctuations that we associate with the small-scale magnetic field (e.g.~in \Eq{rm_sigma2}) might stem from significant spatial variations in the large-scale magnetic fields.

For modelling small-scale components, we study three well-motivated correlation functions (\Sec{mainmethod3}) but such two-point statistics do not account for the non-Gaussian structure of the magnetic fields. It is well known from theory \citep[e.g.][]{zeldovichea1990} and simulations \citep[e.g.][]{SetaEA2020} that the small-scale random magnetic fields are spatially intermittent or highly non-Gaussian. The effect of such random, non-Gaussian structures on $\RM$ statistics is relatively unknown. However, for large $\kpc$ distances to pulsars (where the ratio of the random magnetic field scale to the typical pulsar distance is significantly less than one), the small-scale structure might not matter much for the $\RM$ distribution due to the central limit theorem (assuming point-to-point uncorrelated thermal electron density and magnetic fields).

As discussed in \Sec{mainmethod1}, we assume that the large- and small-scale fields are independent of each other but the tangling of the large-scale field by turbulence could introduce small-scale, random magnetic fluctuations \citep[see Appendix~A in][for a brief discussion]{SetaEA2018}. Thus, they are strictly not independent of each other. Considering such non-trivial connections analytically requires further work but that might introduce corrections to the derived results.

Finally, the pulsar sample is not large enough to differentiate based on specific objects, such as known HII or dense regions, along the line of sight \citep[e.g.~as shown in][]{SobeyEA2021, OckerEA2024}. This might introduce inhomogeneities in the sample that are currently not accounted for by the model. So, some of the overall $\RM$ and $\DM$ variations might be due to these specific line-of-sight effects.

Overall, the currently available data and simplistic modelling allow us to work primarily analytically (except for the numerical minimisation) to study some important parameters of the Galactic thermal electron density and magnetic fields, especially the correlation lengths of the small-scale components, and we aim to explore the impact of these assumptions in our future work.
}

\section{Conclusions} \label{sec:con}
The magneto-ionic medium of the Milky Way, i.e., the thermal electron density in the ionised gas and the magnetic fields, can be divided into large- and small-scale components. This study provides important insights into the structure and dynamics of the magneto-ionic medium of the Milky Way through a comprehensive analysis of pulsar data from the Australia Telescope National Facility pulsar catalogue \citep[][version 2.4.0]{ManchesterEA2005}. In particular, we determine the strength of the large-scale components of the thermal electron density and magnetic fields and scales of their respective small-scale components. Using the dispersion ($\DM$) and rotation ($\RM$) measures of pulsars and analytical models, we have obtained estimates of different properties of thermal electron density and magnetic fields. The main results of the work are given below.

\begin{enumerate}
    \item \textit{{$|\RM|$ vs. $\DM$:}} The analysis showed a strong power-law relationship between the pulsar $|\RM|$ and $\DM$ (\Fig{graph4}~(a) and \Fig{graph5}~(a)). From this, we can conclude that $|\RM| \propto \DM$, which further demonstrates the role of distance; the further away pulsars, accumulate higher $\DM$ and, by extension, $|\RM|$.
    
    \item \textit{Average magnetic field strength vs. distance:} We see a clear lack of correlation between the average parallel magnetic field strength, $\Bpa = 1.232\,\RM / \DM$, with the distance to the pulsar (\Fig{graph4}~(b) and \Fig{graph5}~(b)). It indicates that, for $\RM$, the large-scale magnetic field structure dominates over the sampled distances, with smaller-scale fluctuations averaged out over the path length (further motivating a scale separation of the Galactic magnetic field). However, this does not necessarily imply a uniform large-scale field, and with our method (\Sec{sec:methods}), we determine the large-scale field strength ($\sim 1.2\,\muG$ for distances $\le 20\,\kpc$, see \Tab{results3} for further details) instead of the average magnetic field strength along the line of sight.
    
    \item \textit{Correlation length of small-scale components:} Most importantly, we estimate the scale of the small-scale magnetic field and thermal electron density (see \Tab{results1} and \Tab{results2}). The difference between the correlation length of the magnetic field ($\lb$) and the correlation length of the thermal electron density ($\lne$) is one of the most striking outcomes of this study. We found that $\lb$ is much smaller, within $20$ -- $30$ $\pc$, compared to $\lne$, which is in the range of $250$ -- $300\, \pc$. \rev{Physically, $\lb$ can be seen to be consistent with the theoretical expectations from the turbulent dynamo simulations and $\lne$ with the scales of one or multiple large HII regions (see \Sec{disdiv})}. When studying the magneto-ionic medium or interpreting the extragalactic $\RM$ observations, it is imperative to consider these results because of the very different scales of the thermal electron density and magnetic fields (a factor of $\approx 10$).
    
   \item {\it Impact of the assumed correlation function for the small-scale components:} The choice of correlation function ($\Ca$, $\Cb$, or $\Cc$, \Fig{corfunc}) has practically no effect on the estimated correlation lengths, $\lb$ and $\lne$. Also, large-scale properties of the magneto-ionic medium, such as the mean thermal electron density ($\sim 0.05\,\cm^{{-3}}$) and the large-scale magnetic field strength ($\sim 1.2\,\muG$), remain stable across all correlation functions and are consistent with previous studies.
    
\end{enumerate}

In summary, this study sheds light on the complex structure of the Milky Way's magneto-ionic medium and demonstrates a mathematically concrete way to compute small-scale thermal electron density and magnetic field length scales that factor in the path length and the correlation structure of the random fields. These results can be considered as the starting point for further work that will help sharpen the information about the Galactic magnetic fields and their role in the dynamics and evolution of the Galaxy.

\section{Data Availability}

The data used in the main text of the study are taken from the Australian National Telescope Facility (ATNF) pulsar catalogue \citep[][version 2.4.0, \href{https://www.atnf.csiro.au/research/pulsar/psrcat}{https://www.atnf.csiro.au/research/pulsar/psrcat}]{ManchesterEA2005}. In the appendix, the data from Table~IV.1 of \citet{Ruzmaikin1988} is used. No new data were generated in this research.

\section*{Acknowledgements}
We thank our referee, Anvar Shukurov, for a careful review of this work and for providing insightful comments that helped improve the clarity and rigour of the manuscript. We thank Charlotte Sobey, Yik Ki Ma, Hiep Nguyen, and Naomi M. McClure-Griffiths for useful discussions. AS acknowledges support from the Australian Research Council's Discovery Early Career Researcher Award (DECRA, project~DE250100003) and the Australia-Germany Joint Research Cooperation Scheme of Universities Australia (UA -- DAAD, 2025 -- 2026). This publication is adapted in part from the lead author’s Honours thesis at the Australian National University.

%%%%%%%%%%%%%%%%%%%%%%%%%%%%%%%%%%%%%%%%%%%%%%%%%%



\bibliographystyle{mnras}
\bibliography{pul} % if your bibtex file is called example.bib

\appendix

\section{Derivation of correlation functions}\label{app1}
In this work, we utilised three correlation functions to analyse the small-scale magnetic field and electron density fluctuations within the magneto-ionic medium. 
These three functions are defined and discussed in detail in \Sec{mainmethod3}. In this section, we outline the derivation to compute the contribution of each correlation function to the variance of small-scale fluctuations (see \Sec{mainmethod1} and \Sec{mainmethod2} for the use of this variance).

\subsection{$\Ca$} \label{app1a}
For the exponential correlation function,
\begin{equation}
    \Ca(s) = \Co \exp\left(-{s}/{\ell}\right),
\end{equation}
we want to solve
\begin{equation}
    \varsigma^2 = 2 \int_0^L (L-s) \Ca(s)\, \dd s.
\end{equation}

Substituting $\Ca$ and evaluating the integral (e.g., via integration by parts) gives
\begin{equation}
    \varsigma^2 = 2 \Co \left[ \ell^2 \exp(-L/\ell) - \ell^2 + L\ell \right].
\end{equation}
This provides a compact expression for the variance in terms of $\Co$, $\ell$, and $L$.

\subsection{$\Cb$} \label{app1b}

For $\Cb$, we have
\begin{equation}
    \Cb(s) = \Co \exp\left(-(s/2\ell)^2\right).
\end{equation}

The variance is defined as
\begin{equation}
    \varsigma^2 = 2 \int_0^L (L-s) \Cb(s)\, \dd s
                  = 2 \Co \int_0^L (L-s) \exp\left(-(s/2\ell)^2\right) \dd s.
\end{equation}

Splitting the integral:
\begin{equation}
    \varsigma^2 = 2 \Co \left[ L \int_0^L \exp\left(-(s/2\ell)^2\right) \dd s - \int_0^L s \exp\left(-(s/2\ell)^2\right) \dd s \right].
\end{equation}

Using standard results for the error function and the exponential integral, we obtain
\begin{align}
    \int_0^L \exp\left(-(s/2\ell)^2\right) \dd s &= \ell \sqrt{\pi}\, \operatorname{erf}\left(\frac{L}{2\ell}\right), \\
    \int_0^L s \exp\left(-(s/2\ell)^2\right) \dd s &= 2 \ell^2 \left[1 - \exp\left(-\frac{L^2}{4\ell^2}\right)\right].
\end{align}

Substituting these back gives the final expression for the variance:
\begin{equation}
    \varsigma^2 = 2 \Co \left[ L \ell \sqrt{\pi}\, \operatorname{erf}\left(\frac{L}{2\ell}\right) - 2 \ell^2 \left(1 - \exp\left(-\frac{L^2}{4\ell^2}\right)\right) \right].
\end{equation}

\subsection{$\Cc$} \label{app1c}

For $\Cc$, we have
\begin{equation}
    \Cc(s) = \Co \exp(-s/\ell) \cos(s/\ell).
\end{equation}

The variance is defined as
\begin{equation}
    \varsigma^2 = 2 \int_0^L (L-s) \Cc(s)\, \dd s
                 = 2 \Co \int_0^L (L-s) \exp(-s/\ell) \cos(s/\ell)\, \dd s.
\end{equation}

Using standard integration techniques, this integral evaluates to
\begin{equation}
    \varsigma^2 = \Co \Bigg[ \ell^2 - \exp(-L/\ell) \Big( \ell^2 \cos(L/\ell) + \ell (L \sin(L/\ell) - L \cos(L/\ell)) \Big) \Bigg].
\end{equation}

\section{Ruzmaikin et al. 1998 Data} \label{app2}

\begin{table*} \label{ruztable}
\caption{Parameters for the dataset ($\ndata=27$) taken from Table~IV.1 of \citet{Ruzmaikin1988} using our method (here, like them, we assumed $\gb_0=0^\circ$).}
\begin{tabular}{cccccccc}
\hline 
Method &{$C(s)$} & $\gl_0[\de]$ & $\gb_0[\de]$ & $\co[\cm^{-3}\,\muG$] & $\ell[\pc]$ & $\nea[\cm^{-3}]$ & $B [\muG]$ \\ 
\hline
This work & $\Ca$ & $88 \pm 6 $ & 0 & $-0.05 \pm 0.004$ & $125 $& $0.06 \pm 0.01$ & $1.1 \pm 0.1$\\ 
\citet{Ruzmaikin1988} & $\Ca$ & $99 \pm 12 $ & 0 & $-0.06 \pm 0.002$ & $100-150 $& $0.035 \pm 0.01$ & $2.1 \pm 0.5$\\ 
\hline 
\end{tabular}
\end{table*}

Here, we apply our numerical method to the data in chapter 4 of \citet{Ruzmaikin1988}, which utilised data for 27 pulsars to estimate parameters such as  $\gl_{0}$, $\nea$, and $B$. Note that here we also assume $\gb_{0} = 0^{\circ}$ as they did but in the main text we kept it as a free parameter to be determined from our analysis. The results for both methods are given in \Tab{ruztable}. A key difference in our findings is the value of $\nea$ (higher than their value) and $B$ (lower than their value) and this may also reflect the numerical approach’s ability to capture smaller-scale variations that were probably averaged out in their analytical method. In this work, as described and discussed in the main text, we expand the dataset to include more than 1200 pulsars and multiple correlation functions (they assumed only the exponential decay, $\Ca$) to provide a more comprehensive analysis of the magneto-ionic medium of the Milky Way. 

\rev{
\section{Effect of Non-Gaussianity in RM and DM Distributions}\label{app3}

\begin{table}
\centering
\caption{\rev{Inferred magneto-ionic medium parameters for the $\distdm \leq 20\,\kpc$ sample using different correlation functions ($\Ca$, $\Cb$, $\Cc$) under an assumption of non-Gaussian RM and DM distribution with kurtosis, $\kappa = 5$ (see \Sec{sec:nonGaussian}). Uncertainties represent $1\sigma$ errors. The table demonstrates that the inferred parameters, including $\co$, $\gl_{0}$, $\gb_{0}$, and $\lb$ remain consistent with the Gaussian case (see \Tab{results1}), confirming the robustness of our results against deviations from Gaussianity.}}
\label{tab:appc1}
\begin{tabular}{lcccccc}
\hline
Data Set & $\ndata$ & {$C(s)$} & $c_0$ & {$\gl_0[\degree]$} & {$\gb_0[\degree]$} & $\lb [\pc]$ \\
\hline
$\distdm \leq 20\,\kpc$ & 1290 & $\Ca$ & $0.05 \pm 0.02$ & $192 \pm 9$ & $36 \pm 33$ & $18 \pm 8$ \\
$\distdm \leq 20\,\kpc$ & 1290 & $\Cb$ & $0.05 \pm 0.02$ & $192 \pm 9$ & $36 \pm 33$ & $22 \pm 8$ \\
$\distdm \leq 20\,\kpc$ & 1290 & $\Cc$ & $0.05 \pm 0.02$ & $192 \pm 9$ & $35 \pm 34$ & $18 \pm 8$ \\
\hline
\end{tabular}
\end{table}

\begin{table}\label{residual2}
\centering
\caption{\rev{Similar to \Tab{tab:appc2} but for $\nea$ and $\lne$ using the $\DM$ data. These are also consistent with the results assuming Gaussian $\RM$ and $\DM$ distributions (see \Tab{results2}}).}
\label{tab:appc2}
\begin{tabular}{lcccc}
\hline
Data Set & $\ndata$ & {$C(s)$} & $\nea [\cm^{-3}]$ & $\lne [\pc]$ \\
\hline
$\distdm \leq 20\,\kpc$ &3077 & $\Ca$ & $0.055 \pm 0.001$ & $251 \pm 12$ \\
$\distdm \leq 20\,\kpc$ & 3077 & $\Cb$ & $0.054 \pm 0.001$ & $305 \pm 12$ \\
$\distdm \leq 20\,\kpc$ & 3077 & $\Cc$ & $0.053 \pm 0.001$ & $252 \pm 12$ \\
\hline
\end{tabular}
\end{table}

To assess the sensitivity of our results to the assumption of Gaussian probability density functions (PDFs) for $\RM$ and $\DM$ (via the choice of distribution for $\epsilon_{\RM}$ in \Eq{rm4} and $\eta_{\DM}$ in \Eq{dm4}), we generated non-Gaussian distributions as described in \Sec{sec:nonGaussian}. For both $\RM$ and $\DM$, we tested a range of kurtosis values, $\kappa = 0, 0.1, 1, 2, 5$, for the full sample and for representative subsamples (datasets in \Tab{results1} and \ref{results2}). \Tab{tab:appc1} and \Tab{tab:appc2} illustrates the case with the highest kurtosis, $\kappa = 5$, representing the most extreme departure from Gaussianity.

We find that all the inferred parameters are statistically indistinguishable from those obtained under the Gaussian assumption. Across all tested distributions, the inferred magnetic field and thermal electron density properties remain within the uncertainties of the Gaussian case. These results demonstrate that our conclusions are robust against deviations from Gaussianity, and that the low values of the reduced $\chi^2$ discussed in \Sec{mainmethod4} are unlikely to result solely from the assumption of Gaussian PDFs.
}
%%%%%%%%%%%%%%%%%%%%%%%%%%%%%%%%%%%%%%%%%%%%%%%%%%

% Don't change these lines

%%%%%%%%%%%%%%%%%%%%%%%%%%%%%%%%%%%%%%%%%%%%%%%%%%

% Don't change these lines
\bsp	% typesetting comment
\label{lastpage}
\end{document}